\documentclass[a4paper,11pt]{article}
\usepackage[dvipdfmx]{graphicx}

\usepackage{jheppub} 
\usepackage[T1]{fontenc} 

\title{\boldmath Stokes phenomenon and gravitational particle production --- How to evaluate it in practice}

\author[a,b]{Soichiro Hashiba,}
\author[b]{Yusuke Yamada}

\affiliation[a]{Department of Physics, Graduate School of Science,
The University of Tokyo,\\ Hongo 7-3-1
Bunkyo-ku, Tokyo 113-0033, Japan}
\affiliation[b]{Research Center for the Early Universe (RESCEU), Graduate School of Science,\\ The University of Tokyo, Hongo 7-3-1,
Bunkyo-ku, Tokyo 113-0033, Japan}

\preprint{RESCEU-1/21}

\abstract{We revisit gravitational particle production from the Stokes phenomenon viewpoint, which helps us make a systematic way to understand asymptotic behavior of mode functions in time-dependent background. One of our purposes of this work is to make the method more practical for evaluation of non-perturbative particle production rate. In particular, with several examples of time-dependent backgrounds, we introduce some approximation methods that make the analysis more practical. Specifically, we consider particle production in simple expanding backgrounds, preheating after $R^2$ inflation, and a transition model with smoothly changing mass. As we find several technical issues in analyzing the Stokes phenomenon of each example, we discuss how to simplify the problems while showing the accuracy of analytic estimation under the approximations we make.}

\makeatletter
\gdef\@fpheader{}
\makeatother

\begin{document} 
\maketitle
\section{Introduction}
Particle production in nontrivial backgrounds is one of the most interesting properties of quantum field theory. The Sauter-Schwinger effect~\cite{Sauter:1931zz,Schwinger:1951nm} and the Hawking radiation~\cite{Hawking:1974sw} are well known examples of such effects. A remarkable feature of these effects is that they are non-perturbative with respect to coupling constants. For instance, the Sauter-Schwinger effect cannot be described by the sum of perturbative Feynman diagram processes, which only contain terms proportional to $g^n$ where $g$ is a coupling constant and $n$ is a positive integer. 

Non-perturbative particle production takes place in cosmological contexts as well, and one of the best-known examples is the broad resonance in the preheating after inflation~\cite{Kofman:1994rk,Shtanov:1994ce,Kofman:1997yn}.\footnote{Narrow resonance is not of the kind since it can be expressed as perturbative processes with multiple inflaton particles~\cite{Kofman:1997yn}. In some literature, both narrow and broad resonance are referred to as non-perturbative effects. In our context, only broad resonance regime is a non-perturbative effect, where particle production rate is proportional to a factor $\sim e^{-c/g}$, which is not an analytic function of a coupling constant $g$, and $c$ is typically a quantity depending on momenta and masses of particles. The relation between the broad and narrow resonance regime is similar to non-perturbative production and multi-photon processes in the Sauter-Schwinger effect. }
Although non-perturbative particle production is exponentially suppressed and small in typical cases, repeated events such as the broad resonance in preheating lead to a significant impact on the evolution of universe as well as inflaton dynamics~\cite{Kofman:1997yn}. Non-perturbative particle production would actually play important roles in physics.

Another interesting property of non-perturbative particle production is that heavy particles, which would hardly be produced by perturbative processes, can also be produced. For instance, during the broad resonance in preheating, even if a particle is much heavier than an inflaton particle, a large amount of the heavy particles can be produced, which cannot be the case in perturbative processes by kinematical reasons. The produced heavy particles may become, for instance, unwanted relics which may eventually dominate the universe~\cite{Felder:1999wt}. Therefore, non-perturbative particle production should be taken into account seriously.

How should we evaluate ``particle production'' in nontrivial backgrounds? The definition of a particle is based on solutions of a (free) equation of motion in a given background, which is a second order differential equation. There are in general two independent solutions, which are usually called positive and negative frequency modes, and we expand a field in terms of those solutions associated with creation and annihilation operators, and then define a vacuum state. In a time dependent background, positive and negative frequency solutions defined at early time are related to linear combinations of those defined at late time. Such asymptotic behavior means that the annihilation operator at early time is mixing of the creation and annihilation operators at late time, and the vacuum state defined at early time is no longer one at late time. In this sense, asymptotic behavior of mode functions has one-to-one correspondence with the notion of particle production. Thus, we need mathematical methods to understand the asymptotic behavior of mode functions in a given background, particularly how positive and negative frequency modes mix with each other in late time.

Asymptotic mixing of mode functions can be understood as {\it the Stokes phenomenon} in mathematical language. Such behavior is well described by using the WKB/phase integral method. The relation between particle production and the Stokes phenomenon has been known and applied to the Sauter-Schwinger effect~\cite{Dumlu:2010ua,Dumlu:2010vv,Dabrowski:2016tsx,Taya:2020dco}, the Hawking radiation of blackhole~\cite{Dumlu:2020wvd} and that in de Sitter spacetime~\cite{Kim:2010xm,Kim:2013cka,Dabrowski:2016tsx}, the preheating after inflation~\cite{Enomoto:2020xlf}, particle production in expanding universe~\cite{Kim:2013jca,Li:2019ves} and particle production associated with vacuum decay~\cite{Hashiba:2020rsi}. Understanding of the Stokes phenomenon gives us a systematic way to evaluate the particle production caused by nontrivial background.

The purpose of this paper is to make a more systematic and practical method for evaluation of non-perturbative particle production by focusing on the Stokes phenomenon. In particular, we reconsider the gravitational particle production, which is induced by time varying gravitational backgrounds~\cite{Parker:1969au,Zeldovich:1971mw}, from the Stokes phenomenon viewpoint since curved backgrounds are one of the situations where non-perturbative particle production inevitably takes place. In order to make the method more accessible, we will explicitly show how we can use it with simple examples. Furthermore, we will develop techniques to make {\it analytic} estimates in relatively complicated backgrounds. We will also reconsider whether particle production takes place in simple expanding backgrounds from such a viewpoint.

The analysis based on the Stokes phenomenon has a rigorous mathematical background. However, it might not always be useful in practice if one would like to have an analytic estimation in complicated time-dependent backgrounds. In cosmological and phenomenological models, we often come across with such difficulties. Then, the analysis of the Stokes phenomenon is not so feasible for practical purposes unless approximation methods are established. Therefore, we examine some approximations explicitly and discuss how accurately we can estimate the particle production rate. Besides that, we also discuss how to improve analytic estimation. In our examples, we find that the analysis of the Stokes phenomenon with some approximations can actually gives analytic estimations accurate enough for practical purposes in cosmology as well as phenomenology. 

The rest of this paper is organized as follows. In Sec.~\ref{simple}, we discuss particle production in simple expanding backgrounds. In such backgrounds, no particle production takes place except one nontrivial example. We discuss the nontrivial one in detail, in order to explain basic concepts of the Stokes phenomenon. Section~\ref{oscillation} is devoted to develop techniques for approximation in a complicated background. As a concrete example, we discuss the particle production (preheating) after $R^2$ inflation~\cite{Starobinsky:1980te} where the Hubble parameter oscillates after inflation ends, and curvature induced terms lead to particle production. In Appendix~\ref{appC}, we explain the details of approximations we use in deriving analytic estimations shown in Sec.~\ref{oscillation}. In Sec.~\ref{transition}, we discuss the case that the effective frequency has an infinite number of poles in a complex time plane in addition to turning points, which make the Stokes phenomenon analysis much complicated. We first show the analysis with neglecting poles, and then improve it by taking into account a pole contribution. To the best of our knowledge, the Stokes line analysis with poles has not been shown explicitly in the context of particle production.\footnote{\cite{Kutlin} has discussed the contribution from poles in the context of quantum tunneling. In~\cite{Kim:2013jca}, the author takes pole contributions into account in models of nontrivial background geometries, but the formalism seems different from ours. For example, the notion of Stokes lines is not discussed in~\cite{Kim:2013jca}. In this sense, our approach would be complementary to that in~\cite{Kim:2013jca}.} Our analysis would clarify the importance of a pole contribution in order to improve the estimate, or said oppositely, it would tell us when we may neglect such a contribution. The detailed analysis of a pole contribution is given in Appendix~\ref{appB}. Finally, we conclude in Sec.~\ref{conclusion}. For readers who are not familiar with the Stokes phenomenon, we give a brief review of it in Appendix~\ref{appA}, which contains minimal information necessary for this paper.

Throughout this paper, we will take the natural units $c=\hbar=1$.

\section{Particle production in simple expanding Universe models: warm up}\label{simple}
We consider a massive scalar $\phi$ coupled to the Ricci scalar $R$ in expanding backgrounds and the action of the scalar field is given by
\begin{equation}
    S =\int d^4x\mathcal{L}= -\frac12\int d^4x\sqrt{-g}\left[ g^{\mu\nu} \partial_\mu\phi \partial_\nu\phi + m^2\phi^2 + \xi R\phi^2\right],\label{scalar}
\end{equation}
where $m$ and $\xi$ are a scalar mass and a coupling constant, respectively. In a flat expanding background $ds^2=a^2(\eta)(-d \eta^2+d{\bf x}^2)$, where $a$ is a scale factor and $\eta$ is a conformal time $d\eta=dt/a$, the action of the scalar field is reduced to
\begin{equation}
    S = \frac12\int d\eta d^3{\bf x}\left[{\chi}'^2-(\partial_i\chi)^2-\left(m^2a^2+(6\xi-1)\frac{a''}{a}\right)\chi^2\right],
\end{equation}
where a prime denotes the derivative with respect to the conformal time, and we have introduced a rescaled scalar $\chi\equiv a\phi$. The canonically quantized scalar $\chi$ can be expanded as follows:
\begin{equation}
    \chi({\bf x},\eta)=\int\frac{d^3{\bf k}}{(2\pi)^{3/2}}\left(\hat{a}_{\bf k}v_k(\eta)e^{{\rm i}{\bf k}\cdot{\bf x}}+\hat{a}_{\bf k}^\dagger \bar{v}_k(\eta)e^{-{\rm i}{\bf k}\cdot{\bf x}}\right),
\end{equation}
where $\hat{a}_{\bf k} (\hat{a}_{\bf k}^\dagger)$ denotes the annihilation (creation) operator that obeys the canonical commutation relation $[\hat{a}_{\bf k},\hat{a}^\dagger_{{\bf k}'}]=\delta({\bf k}-{\bf k}')$, $v_k(\eta)$ and $\bar{v}_k(\eta)$ denote the mode function and its complex conjugate, respectively. The equation of motion of the scalar leads to
\begin{equation}
    v_k''+\left(k^2+m^2a^2+(6\xi-1)\frac{a''}{a}\right)v_k=0.\label{meq}
\end{equation}
We see that the effective frequency of the mode function is time-dependent, which implies that particle creation inevitably takes place due to the background $a(\eta)$.

In order to discuss the particle production in time-dependent backgrounds, we will use the WKB/phase integral method, which would give a universal way to analyze non-perturbative particle creation. For readers who are not familiar with this method and the Stokes phenomenon, we briefly review these mathematical backgrounds in Appendix~\ref{appA}. In the following, we give a qualitative review about the Stokes phenomenon. We consider the following two basis functions
\begin{equation}
    f_{\pm}(\eta)=\frac{1}{\sqrt{2\omega_k(\eta)}}e^{\pm{\rm i}\int^\eta d\eta' \omega_k(\eta')},
\end{equation}
where
\begin{equation}
    \omega_k^2=k^2+m^2a^2+(6\xi-1)\frac{a''}{a}
\end{equation}
is the effective frequency. These basis functions can be the actual solution of \eqref{meq} in the adiabatic limit $\omega_k\to{\rm const}$. In a sufficiently adiabatic region, $\omega'_k/\omega_k^2\ll1$, the mode function would be well approximated by a linear combination of them as
\begin{equation}
v_{k}(\eta)=\alpha_k f_- + \beta_k f_+,
\end{equation}
where $\alpha_k$ and $\beta_k$ are the Bogoliubov coefficients, which are normalized as $|\alpha_k|^2 - |\beta_k|^2 = 1$ for a scalar field. In particular, the adiabatic vacuum mode function is given by $\alpha_k(\eta_0)=1, \beta_k(\eta_0)=0$ if $\omega_k\to {\rm const.}$ as $\eta\to \eta_0$. For a general $\eta_0$, $\alpha_k(\eta_0)=1$ and $\beta_k(\eta_0)=0$ would be referred to as the instantaneous Hamiltonian diagonalization. The choice of these conditions on $\alpha_k$ and $\beta_k$ corresponds to the ``vacuum'' state one takes. In the following, we take $\alpha_k=1,\beta_k=0$ at some initial time $\eta = \eta_0$ as an initial condition.

\begin{figure}[htbp]
\centering
\includegraphics[width=.90\textwidth]{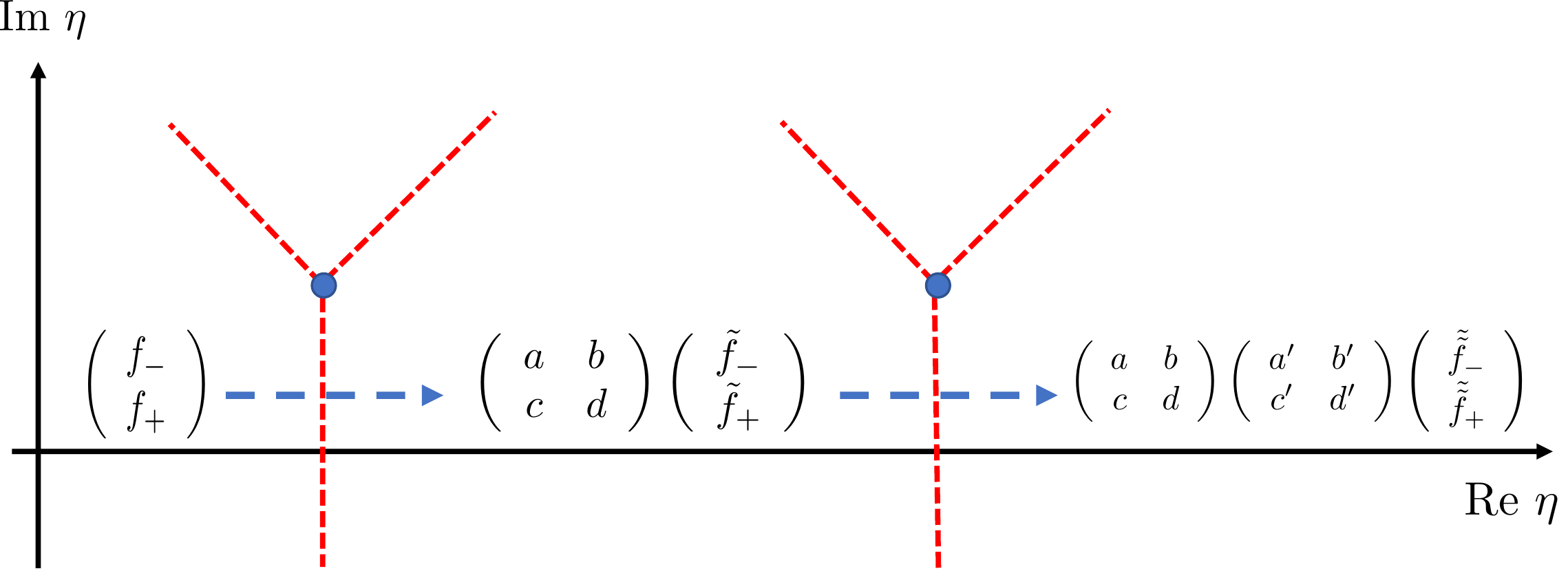}
\caption{\label{fig:connection} Schematic picture of the Stokes phenomenon in complex $\eta$-plane. The vertical and horizontal axis correspond to imaginary and real $\eta$-axis, respectively. The blue blobs correspond to turning points, where the effective frequency vanishes $\omega_k^2=0$. From each turning point, three Stokes lines (red dashed lines) emanate, which separate the region where the WKB/phase integral solutions are well defined. The locally defined WKB/phase integral solutions can be connected to other region, which in general mixes the positive and negative frequency modes. We interpret such connection as particle production. When multiple Stokes lines exist, we need to consider connection problems at each crossing.}
\end{figure}

The WKB solutions or more generally, the phase integral solutions are useful to understand {\it the Stokes phenomenon}: WKB/phase integral solutions are locally defined asymptotic series, and not globally. In this sense, complex $\eta$-plane is separated in several regions where basis functions are locally defined. The border lines for such division are given by so-called Stokes lines, which emanate from turning points $\eta=\eta_c$ satisfying $\omega_k(\eta_c)=0$. In order to follow the behavior of the mode functions, one needs to consider the connection problems of basis functions defined in different regions. In general, the connection is expressed as (see Fig.~\ref{fig:connection})
\begin{equation}
    \left(\begin{array}{c}f_-(\eta)\\ f_+(\eta)\end{array}\right)=\left(\begin{array}{cc}
        a & b  \\
        c & d
    \end{array}\right)\left(\begin{array}{c}\tilde{f}_-(\eta)\\ \tilde{f}_+(\eta)\end{array}\right),\label{connection}
\end{equation}
where $a,b,c,d$ are constants satisfying $ad-bc=1$ following from the Wronskian condition. Here, $f_\pm$ and $\tilde{f}_{\pm}$ are the basis functions defined in different regions in complex $\eta$-plane, which are separated by Stokes lines. We can interpret such mixing of the positive and negative frequency solutions as a (non-perturbative) particle production ``event''.\footnote{In some situation, this event may correspond to rather particle ``annihilation''. However, it is not the case as far as we start from the adiabatic vacuum condition. Particle ``annihilation'' takes place when we consider the interference of several events~\cite{Kofman:1997yn}.} In particular, if $b\neq0$ and one takes the adiabatic vacuum mode function as an initial condition, the change can be regarded as particle production, where the number density of the produced particle is given by $n_k=|\beta_k|^2=|b|^2$. 
We should emphasize that although we will discuss ``particle production'' throughout this paper, in most models (except a model discussed in Sec.~\ref{transition}), the notion of particle is not well-defined in a strict sense, since backgrounds do not become static after crossing Stokes lines. Nevertheless, in order to check the accuracy of our analytic formula for connection matrix elements, we will compare $n_k=|\beta_k|^2$ with numerical solutions of mode equations for illustrative purposes. 

We also note that if there are multiple Stokes lines that cross the real $\eta$-axis, one needs to consider a connection problem at each Stokes line crossing (see Fig.~\ref{fig:connection}). Then, we multiply a connection matrix as in \eqref{connection}. Throughout this paper, however, we will only consider particle production at one event and will not treat multiple events. The multiple events may lead to interesting effects such as interference, see e.g.~\cite{Dumlu:2010ua,Dumlu:2010vv,Taya:2020dco}.

In the following, we will discuss the structure of Stokes lines in some simple expanding backgrounds and evaluate produced particle number density.

\subsection{Simple expanding universe}
Let us consider the general (isotropic) expanding background where the energy density $\rho$ and the pressure $p$ of the Universe satisfy the equation of state $p=w\rho$. Assuming that $a(t_0)=1$ where $t_0$ is the initial physical time and $\eta=0$ is the initial conformal time without loss of generality,\footnote{For a conformally coupled massive scalar in this background, the time $a(t_0)=0$ could be an initial time since the scalar field equation of motion is smooth at the time. Nevertheless, since $a(t_0)=0$ is an initial singularity and particle picture is no longer well-defined, we will take $a(t_0)=1$ as the initial condition.\label{f4}} the scale factor is given by
\begin{equation}
    a(\eta)=\left(H_i\eta+1\right)^{\frac{2}{1+3w}},
\end{equation}
where $H_i=\frac{1+3w}{2}H(t_0)$. We should note that some of ``particle production'' discussed in this section are not really particle production associated with expansion of the Universe, but just rephrasing of the known behavior of a scalar field in expanding backgrounds. Nevertheless, it would be useful to understand the dynamics of a scalar field from the Stokes phenomenon viewpoint. The relevant quantities for our discussion are given by
\begin{align}
   \frac{ a''}{a}=&\frac{2(1-3w)H_i^2}{(1+3w)^2(H_i\eta+1)^2},\\
   a^2=&(H_i\eta+1)^{\frac{4}{1+3w}},
\end{align}
and these yield the effective frequency:
\begin{equation}
     \omega_k^2=k^2+m^2(H_i\eta+1)^{\frac{4}{1+3w}}+(6\xi-1)\frac{2(1-3w)H_i^2}{(1+3w)^2(H_i\eta+1)^2}.
\end{equation}
For concreteness, we consider the case $w=0, 1/3, 1$ which corresponds to the matter, radiation, and kinetic energy dominated era, respectively. 

\subsection{Nontrivial example: conformally coupled massive scalar in matter dominated Universe}
\begin{figure}[htbp]
\centering
\includegraphics[width=.60\textwidth]{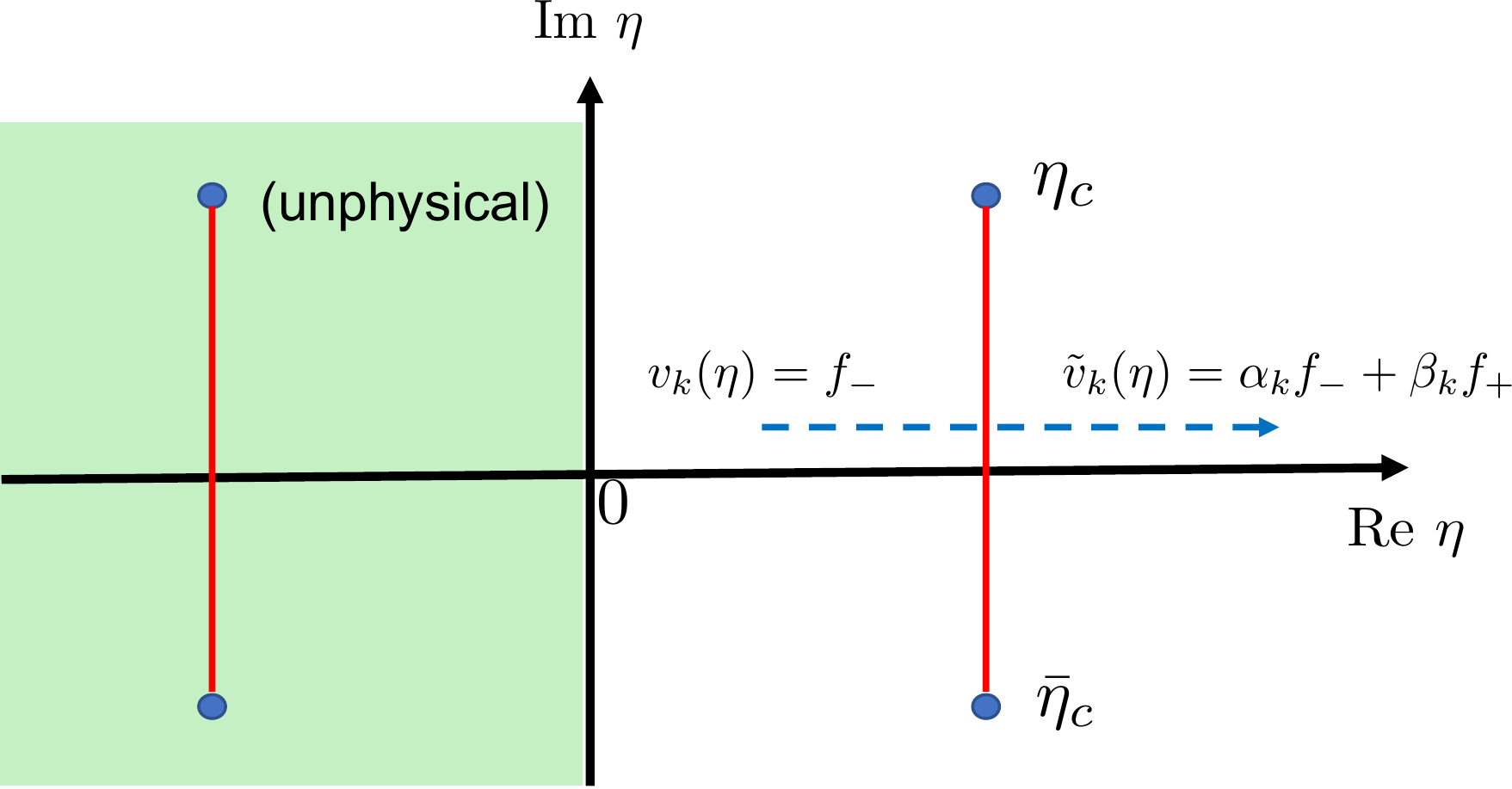}
\caption{\label{fig:confmass} The structure of Stokes lines (more precisely, Stokes segments, see Appendix~\ref{appA}) for the effective frequency~\eqref{confmass} in complex $\eta$-plane. We only consider the physical region ${\rm Re}\:\eta>0$, and the green shaded region corresponds to unphysical one. The blue dots and the red lines denote turning points and Stokes lines, respectively. We consider the connection from ${\rm Re}\:\eta= 0$ to $\eta\gg {\rm Re}\:\eta_c$. We assume the adiabatic vacuum as the initial state, but after crossing the Stokes line, the mode function becomes a linear combination of the positive and negative frequency modes, namely particle production takes place. Although the turning points appear also in the unphysical region, we do not need to consider crossing the Stokes line emanating from them. We also note that we have omitted the Stokes lines not crossing the real axis.}
\end{figure}

For a conformally coupled scalar field in matter dominated Universe, $w=0$, $\xi=\frac16$, we have the effective frequency
\begin{equation}
    \omega_k^2 = k^2+m^2(H_i\eta+1)^4. \label{confmass}
\end{equation}
In this case, turning points $\eta_c$ satisfying $\omega_k(\eta_c)=0$ appear at 
\begin{align}
    H_i\eta_c+1 = -\sqrt{\frac{k}{2m}}(1\pm {\rm i}),\ \sqrt{\frac{k}{2m}}(1\pm {\rm i}).
\end{align}
A schematic picture of the Stokes lines in this model is shown in Fig.~\ref{fig:confmass}. Since $\eta=0$ is the initial time, Stokes line crossing the real time axis before this time is unphysical. On the other hand, there is a Stokes line connecting the turning point $H_i\eta_c=\sqrt{\frac{k}{2m}}(1\pm {\rm i})-1$ and its conjugate, which cross the real axis after $\eta=0$ for $k>2m$.\footnote{We should also note that the turning points have positive real parts only for the modes $k>2m$. However, the boundary value $k=2m$ itself does not have any specific physical meanings. Rather, this value is determined by our initial condition $a(t_0)=1$. As quoted in Footnote~\ref{f4}, we could take $a(t_0)=0$ or equivalently, $\eta=-1/H_{i}$ as an initial time. If we took $\eta=-1/H_{i}$ as an initial time, there are two turning points that have real parts larger than ${\rm Re}\:\eta_c>-1/H_i$ for all modes $k>0$. Namely, all the modes cross a Stokes segment connecting those turning points. However, defining an adiabatic particle state at initial singularity would not make sense.}

\begin{figure}[htbp]
\centering
\includegraphics[width=.70\textwidth]{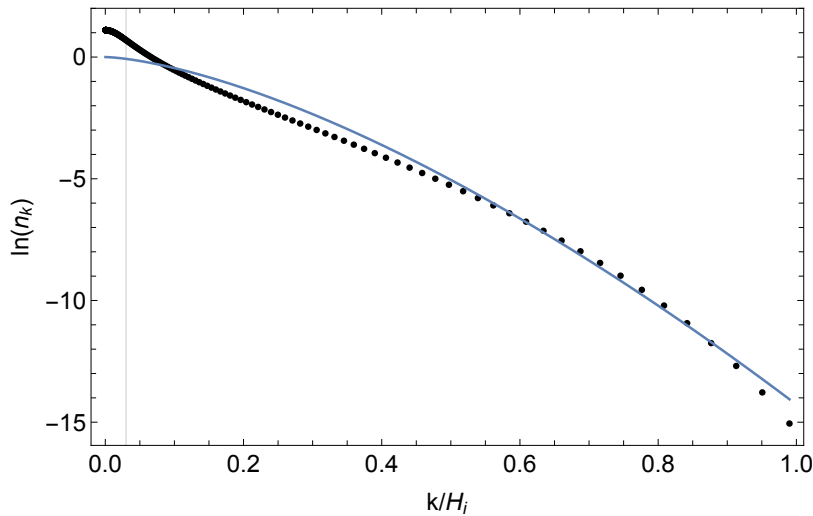}
\caption{\label{fig:MDm003} Comparison between a numerical calculation (black dots) and an analytic estimate~\eqref{nkMD} (blue line). The vertical gray thin line denotes $k=2m$. Here, we take $m = 0.03H_i$.}
\end{figure}

We take the adiabatic vacuum condition near $\eta=0$, namely, the initial mode function is \begin{equation}
    v_k(\eta)=\frac{1}{\sqrt{2\omega_k(\eta)}}e^{-{\rm i}\int^{\eta}_0 d\eta'\omega_k(\eta')}.
\end{equation}
Then, we consider the behavior of the mode function in later time $\eta\gg {\rm Re}\ \eta_c$. After the Stokes line crossing, the initial mode function is analytically continued to a linear combination of the positive and negative frequency modes,
\begin{equation}
    v_k(\eta)=\frac{1}{\sqrt{2\omega_k(\eta)}}e^{-{\rm i}\int^{\eta}_0 d\eta'\omega_k(\eta')}\to \frac{\alpha_k}{\sqrt{2\omega_k(\eta)}}e^{-{\rm i}\int^{\eta}_0 d\eta'\omega_k(\eta')}+\frac{\beta_k}{\sqrt{2\omega_k(\eta)}}e^{+{\rm i}\int^{\eta}_0 d\eta'\omega_k(\eta')},
\end{equation}
where $\alpha_k$ and $\beta_k$ are Bogoliubov coefficients. We need to evaluate in particular the coefficient of the negative frequency mode, $\beta_k$. As discussed in Appendix~\ref{appA}, the Bogoliubov coefficient at the Stokes line crossing is approximately given by
\begin{equation}
   \beta_k={\rm i}\exp\left[{\rm i}\int _{\bar{\eta}_c}^{\eta_c}d\eta\omega_k\right].
\end{equation}
One can simply perform this integration and find the resultant particle number density to be
\begin{equation}
    n_k=|\beta_k|^2=\exp\left[-\frac{(\Gamma(\frac14))^2}{3\sqrt{\pi}}\frac{m}{H_i}\left(\frac{k}{m}\right)^{\frac32}\right], \label{nkMD}
\end{equation}
where $\Gamma(x)$ is Gamma function. Figure~\ref{fig:MDm003} shows the comparison between our analytic estimate~\eqref{nkMD} and numerical results, which shows their agreement up to ${\cal O}(1)$ factor. Note that this particle production can take place only for modes with $k>2m$ for consistency as discussed above. Taking into account it, the total number density is \begin{equation}
    n=4\pi\int_{k=2m}^\infty dkk^2 n_k=\frac{8\pi m^3(1+2\sqrt{2}c)e^{-2\sqrt{2}c}}{3c^2},
\end{equation}
where $c=\frac{\Gamma(1/4)^2m}{3\sqrt{\pi}H_i}$. 

We would like to give some comments on the comparison between numerical results and our analytic estimate. As shown in Fig.~\ref{fig:MDm003}, the numerical result shows non-vanishing particle number for the modes $k<2m$, which seems inconsistent with our expectations. In the numerical calculation, we have taken the adiabatic initial condition at $\eta=0$. A possible explanation of such a discrepancy is that the low momentum modes are not adiabatic around $\eta=0$. Therefore, our semi-classical analysis might not be appropriate.

We finally note that, although the total ``particle number'' given by momentum integration of the square of the Bogoliubov coefficient asymptotes to a fixed value,  the ``particle production'' discussed here should not be interpreted as production of real ``particles'' in a strict sense, since the effective mass of the particle continuously grows during this era, and one particle state is not well defined. Hence, the Bogoliubov coefficients derived above should be interpreted as just evolution of the field. Nevertheless, the Bogoliubov coefficients, or more precisely, the connection matrix would be inherited by the asymptotically defined particle with stationary frequency.\footnote{If the expansion terminates around some point, it would mean that scale factor changes its behavior there, which changes the scale factor from $a^2(\eta)=(H_i\eta+1)^4$ to some different form. Such behavior of the scale factor may lead to additional turning points and Stokes lines. Even in such a case, if the Stokes line we discussed and the additional ones are well separated, the connection problem discussed here is not affected, namely the connection matrix is preserved. Then, we need to discuss the connection problem at the additional Stokes line crossing. Here we will not discuss such an issue since it requires to fix the scale factor evolution, namely the history of the Universe.}

\subsection{Other cases: no particle production}
We discuss whether non-perturbative particle production can take place in other expanding backgrounds. The answer is negative, and we find no particle production in simple setups we discuss below. The Stokes phenomena discussed in the following are reinterpretations of known dynamics of mode function in each case.

\subsection*{Minimally coupled massless scalar in matter dominated Universe}
Let us consider a simple case with minimally coupled massless scalar $m=0,\xi=0$ in matter dominated era $w=0$. In this case, the effective frequency is $\omega_k^2=k^2-2H_i^2(H_i\eta+1)^{-2}$ and small $k$ modes seem to have instability $\omega_k^2<0$. Such modes eventually find the turning point $H_i\eta_c+1=\frac{\sqrt{2}H_i}{k}$. Although this is a Stokes phenomenon, we cannot interpret it as particle production simply because we cannot define particle at the initial time $\eta=0$. We notice that such an ``instability'' appears for the modes outside of the horizon, which is reasonable since we cannot have a notion of particle for such modes. We also find that the turning point depends on $k$ and its value becomes larger for smaller $k$. We may interpret the turning point to be the time around which the mode enters the horizon. Only after passing the turning point $H_i\eta_c+1=\frac{\sqrt{2}H_i}{k}$, $\omega_k$ becomes real and we are able to introduce the notion of particles. Similar behavior can be found in the kination era as we will see below. We also note that in this case, one can solve the mode equation exactly, and can check the behavior of the mode function.

\subsection*{In radiation dominated Universe}
Let us consider the radiation dominated era $w=1/3$, where the effective frequency is given by
\begin{equation}
    \omega_k^2=k^2+m^2(H_i\eta+1)^2.
\end{equation}
In this case, the turning point is $H_i\eta_c+1=\pm {\rm i}\frac{k}{m}$. We find that the real part of the turning point is $-1/H_i$, where the scale factor is $a(\eta)=0$.\footnote{Strictly speaking, a location of the turning points are not crucial, and the place of particle production is determined by the point where Stokes line crosses the real axis. However, in this case, the Stokes line connecting two turning points are straight line, and therefore, the particle production associated with the turning points are unphysical.} If the radiation dominated era starts from $a=1$, this point will not be crossed and therefore we may identify this point to be unphysical turning points. Because of vanishing curvature contribution, the presence of the non-minimal coupling does not change the structure of frequency $\omega_k$. From this observation, we conclude that there is no gravitational particle production in radiation dominated era.

\subsection*{Kinetic energy dominated (kination) Universe}
In kination era, the equation of state parameter is $w=1$, and we have
\begin{equation}
    \omega_k^2=k^2+m^2(H_i\eta+1)-\frac{(6\xi-1)}{4}H_i^2(H_i\eta+1)^{-2}.
\end{equation}

For conformally coupled case $\xi=1/6$, a turning point appears at $H_i\eta+1=-k^2/m^2$ which is unphysical, and there is no particle production. The next simple example is the case of $\xi<1/6$ with $m=0$. In this case, again, the turning point appears at unphysical region, and no particle production takes place. 

For $\xi>1/6$ with $m=0$, we find a turning point at $H_i\eta_c+1=\frac{\sqrt{6\xi-1}}{2k}$ on the real $\eta$-axis. We notice that, until that time, the square of the frequency becomes negative, which means instability for the mode. This is known as spinodal instability in kination era~\cite{Nakama:2018gll}. This ``instability'' is similar to that in matter dominated era discussed above. As is the case of the ``tachyonic mode'' in matter dominated era, until the time reaches the turning point, such a mode cannot be interpreted as a particle since its frequency is not real for $H_i\eta+1<\frac{\sqrt{6\xi-1}}{2k}$. Therefore, it is more appropriate to understand this turning point to be the time when the corresponding mode reenter the horizon. Only after the reentering, we can define a particle, which is physically reasonable. If the scalar had quantum fluctuation prior to the kination era, its amplitude can be conserved until horizon reenter. After this point, the mode start to oscillate around the minimum, which may eventually dominate the universe since the kination energy decays as $a^{-6}$. Such a mechanism has been applied to overcome the graviton overproduction problem in \cite{Nakama:2018gll}.

\subsection*{Comments on the simple expansion models}
As we have seen above, most of simple expanding Universe models would not lead to non-perturbative particle production. One of the reasons for the absence of the particle production is that there are only few turning points. As we will see in the following section, more nontrivial time-dependent background will lead to non-perturbative particle production. One may expect it simply because more complicated time dependence may lead to more turning points and Stokes lines, namely particle production ``events''. Such a naive expectation is actually true in the following examples.

\section{Hubble oscillating phase in $R^2$ model: imitation method}\label{oscillation}
In this section, we discuss scalar particle production (preheating) in $R^2$ inflation model~\cite{Starobinsky:1980te}, which is supported by CMB observations~\cite{Akrami:2018odb}. However, if one wants to analyze this model exactly, the Hubble parameter becomes very complicated and accordingly it is hard to straightforwardly analyze Stokes lines and to make analytic estimates. Therefore, in practice, some approximations are requisite for an analytic estimate of particle production rate. The main purpose of this section is to show how the approximation of the background and the Stokes phenomenon analysis based on it affect the accuracy of analytic estimates.

In this model, one can use the Jordan frame as well as the Einstein frame by introducing a scalar degree of freedom called a scalaron, but for our purpose, we use the Jordan frame since all dynamics can be written in terms of gravity.\footnote{Since we are assuming non-minimal coupling, there are mixing terms between scalaron and the scalar $\phi$ besides the terms coming from metric. In Jordan frame, the scalaron mixing terms are included in gravitational coupling.} After inflationary phase, the Hubble parameter and the Ricci scalar starts to oscillate, which corresponds to scalaron oscillation in the Einstein frame. We focus especially on the case with a sizable non-minimal coupling to Ricci scalar, where the particle production in this phase becomes efficient. For the gravitational part of the action, we assume
\begin{equation}
 S_g = \frac{M_{\rm pl}^2}{2}\int d^4x\sqrt{-g}\left(R+\frac{R^2}{6M^2}\right),
\end{equation}
where $M$ denotes a mass parameter corresponding to the scalaron mass in Einstein frame. The equations of motion for Hubble parameter and Ricci scalar are (see e.g.~\cite{DeFelice:2010aj})
\begin{align}
   \ddot{H}-\frac{\dot{H}^2}{2H}+\frac12 M^2 H &= -3H\dot{H},\\
    \ddot{\hat{R}}+\left(M^2-\frac34 H^2-\frac32 \dot{H}\right)\hat{R} &= 0,
\end{align}
where $\hat{R}\equiv a^{3/2}R$. Here we have taken spacetime metric to be $ds^2=-dt^2+a^2(t)d{\bf x}^2$ and the Hubble parameter is $H=\dot{a}/a$. Inflation ends around $t=t_{\rm os}$ where $\dot{H}=-M^2/6$, which is the initial condition of the oscillating phase. With such an initial condition and the equation of motion, we find
\begin{equation}
    H(t)=\left(\frac{3}{M}+\frac34 (t-t_{\rm os})+\frac{3}{4M}\sin M(t-t_{\rm os})\right)^{-1}\cos^2\left(\frac{M}{2}(t-t_{\rm os})\right).\label{Hosc}
\end{equation}
Hereafter, we will take $t_{\rm os} = 0$ without loss of generality. 

Let us discuss particle production in this background. The action of the scalar $\phi$ is given in \eqref{scalar}. So far, we have used conformal time to analyze the particle production.\footnote{The reason why we have used the conformal time in previous discussion is that such coordinate simplifies analysis in cases where a scale factor is given by a simple power law. There, conformal time expression avoids some branch cut appearing in effective frequency $\omega_k$, which leads to unnecessary complication.}  However, at this stage, physical time coordinate $t$ is more appropriate for our analysis since rewriting time evolution with conformal time makes various quantity more complicated. With the flat expanding coordinate $ds^2=-dt^2+a^2(t)d{\bf x}^2$, the canonically normalized scalar field action is given by
\begin{equation}
    S=\frac12 \int dt dx^3\left[\dot{\psi}^2-\frac{1}{a^2}(\partial_i\psi)^2-\left(m^2+\frac32(4\xi-1)\dot{H}+\frac34(16\xi-3)H^2\right)\psi^2\right],
\end{equation}
where $\psi \equiv a^{3/2}\phi$. Accordingly, we quantize the rescaled scalar field as 
\begin{equation}
    \psi=\int\frac{d^3{\bf k}}{(2\pi)^{3/2}}\left(\hat{a}_{\bf k}u_k(t)e^{{\rm i}{\bf k}\cdot{\bf x}}+\hat{a}_{\bf k}^\dagger \bar{u}_k(t)e^{-{\rm i}{\bf k}\cdot{\bf x}}\right).
\end{equation}
The mode equation for $u_k(t)$ becomes
\begin{align}
    \ddot{u}_k(t)+\Omega_k^2u_k(t)=0, \label{pureeom}
\end{align}
where the effective frequency is\footnote{Equation \eqref{Omegak} has the curvature induced term even when $\phi$ is ``conformally'' coupled ($\xi = 1/6$). This is due to our choice of frame. If we take the comoving frame and rescale the field as $\chi = a\phi$, the curvature-dependent term vanishes at $\xi = 1/6$. However, this difference is negligible when we consider a large $\xi$ case.}
\begin{equation}
\Omega_k^2=\frac{k^2}{a^2}+m^2+6\left(\xi-\frac{1}{4}\right)\dot{H}+12\left(\xi-\frac{3}{16}\right)H^2.\label{Omegak}
\end{equation}

\begin{figure}[htbp]
\centering
\includegraphics[width=.84\textwidth]{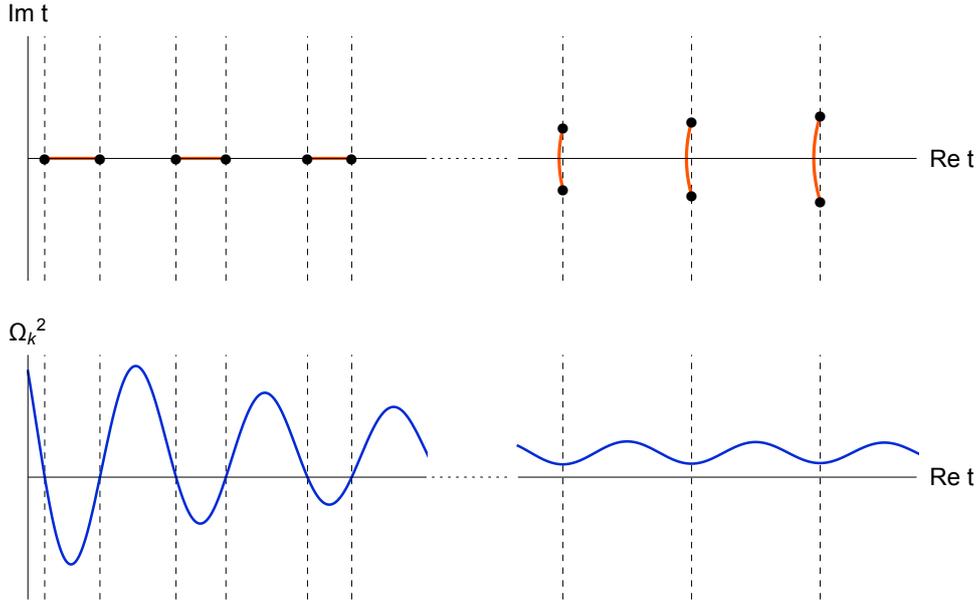}
\caption{\label{fig:osc_SL} Lower panels: Damped oscillation of effective frequency. Upper panels: Corresponding turning points (black dots) and Stokes segments (red lines) in complex time plane. Dashed lines show zeros (left panels) and minima (right panels) of the effective frequency. When the amplitude of the effective frequency is so large that it can become tachyonic, its Stokes segments appear along real time axis.}
\end{figure}

The effective frequency $\Omega_k$ exhibits damped oscillation as shown in Fig.~\ref{fig:osc_SL}. In earlier time when the amplitude of the effective frequency is still so large that it can become tachyonic, Stokes segments lie along the real time axis where the effective frequency is tachyonic, and become shorter and shorter for later oscillation (left panels of Fig.~\ref{fig:osc_SL}). Eventually the effective frequency becomes positive definite as the oscillation is damped, and then, Stokes segments become almost perpendicular to the real time axis, and cross it near the minima of the effective frequency. In this regime, the length of the segment becomes longer and longer as the oscillation is more damped (right panels of Fig.~\ref{fig:osc_SL}). Each Stokes line crossing corresponds to particle production ``event'', and the amount of produced particles at each event can be evaluated as explained in Appendix~\ref{appA1} and~\ref{appA2}, respectively. The physical interpretation of this mathematical structure is that particles are produced mostly when they become tachyonic or lightest as expected. Note that this model has a similar structure as that discussed in~\cite{Dufaux:2006ee}, where preheating from trilinear interaction is discussed and the tachyonic regime dominates particle production.

\begin{figure}[htbp]
\centering
\includegraphics[width=.50\textwidth]{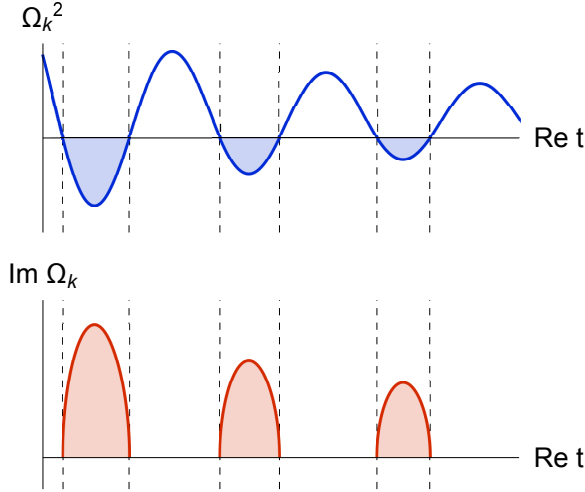}
\caption{\label{fig:osc_tac} The time evolution of the squared value (upper panel) and the imaginary part (lower panel) of the effective frequency $\Omega_k$. The right-hand side of~\eqref{hombuntac} corresponds to the area of the red shaded region.}
\end{figure}

\begin{figure}[htbp]
\centering
\includegraphics[width=.64\textwidth]{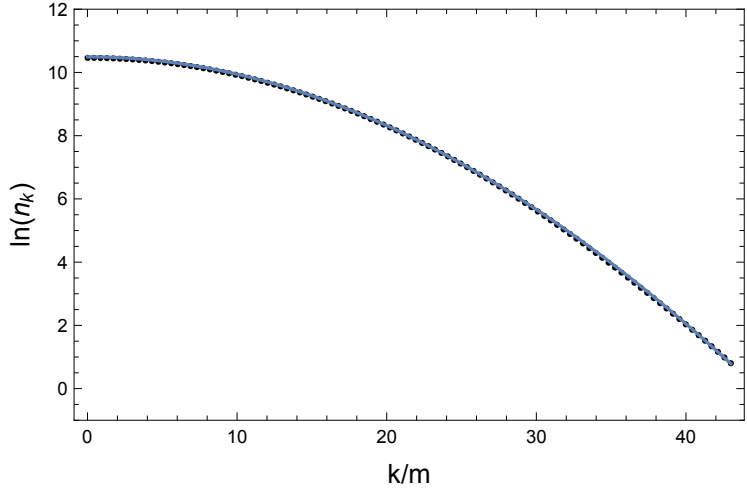}
\caption{\label{fig:osc_exact} The particle number produced by the first oscillation of the Ricci scalar. Black dots and the blue solid line depict the numerical result obtained by solving the equation of motion with the effective frequency~\eqref{Omegak} and the approximation~\eqref{eqappA1}, respectively. In this graph, we take $M=15.0m, \xi=10.0$. This shows that our approximation formula~\eqref{eqappA1} actually well describes the particle production rate.}
\end{figure}

Since the particle production is much more efficient in the tachyonic regime, in the following, we will focus only on that regime. The WKB/phase integral solution is
\begin{equation}
    u_k(t)=\frac{\alpha_k(t)}{\sqrt{2\Omega_k(t)}}e^{-{\rm i}\int^t dt' \Omega_k(t')}+\frac{\beta_k(t)}{\sqrt{2\Omega_k(t)}}e^{+{\rm i}\int^t dt' \Omega_k(t')}.
\end{equation}
The Bogoliubov coefficient $\beta_k$ is important to evaluate the particle number, and in Appendix~\ref{appA1} we show how the Bogoliubov coefficient $\beta_k$ is calculated after passing one Stokes segment in tachyonic regime. With the approximate formula~\eqref{eqappA1}, $\beta_k$ is evaluated by the time integral of the effective frequency in the tachyonic domain:
\begin{equation}
    |\beta_k| = \exp\left(\int_{t_{c,2n}}^{t_{c,2n+1}} |\Omega_k| dt\right), \label{hombuntac}
\end{equation}
where $t_{c,2n}$ and $t_{c,2n+1}$ ($n=0,1,2,\cdots$) denote the $(n+1)$-th pair of turning points. Therefore, particle number can be simply given by integrating the effective frequency~\eqref{Omegak} along the tacyonic Stokes segments, in other words, the area of the red shaded region in the lower panel of Fig.~\ref{fig:osc_tac}. In Fig.~\ref{fig:osc_exact}, we compare the numerically integrated Bogoliubov coefficient~\eqref{hombuntac} and the particle number evaluated with the numerical solution of the mode equation~\eqref{pureeom}, and confirm that the evaluation of~\eqref{hombuntac} gives the correct particle number. Hence, we will discuss how to derive the analytic expression of~\eqref{hombuntac} despite the complication of the effective frequency.

\begin{figure}[htbp]
\centering
\includegraphics[width=.64\textwidth]{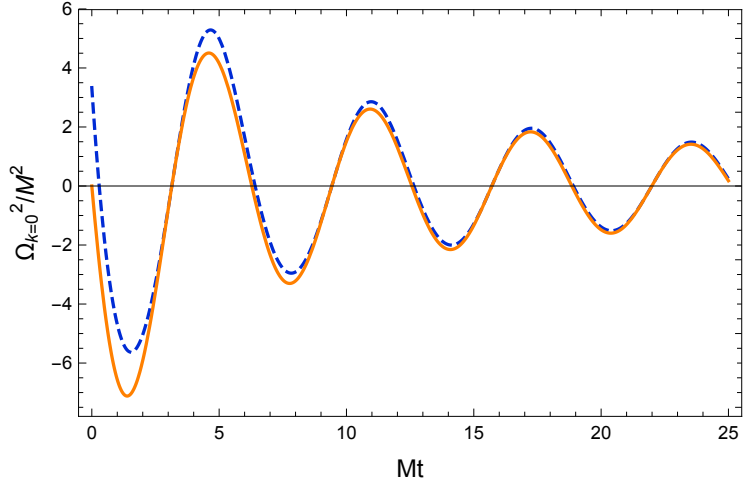}
\caption{\label{fig:oscR} Comparison between the curvature induced terms of the exact effective frequency derived from the Hubble parameter~\eqref{Hosc} (blue dashed line) and one of the approximated form~\eqref{Omegaapprox} (orange solid line). In this graph, we take $k=0, m=0, \xi = 10.0$. The approximation becomes better as $Mt$ becomes larger.}
\end{figure}

Let us discuss how to make an analytic estimate for the particle production. Since the Hubble parameter given in \eqref{Hosc} is too complicated, it seems difficult to find turning points and to integrate the effective frequency along Stokes lines analytically, although numerical evaluation is possible as shown in~Fig.~\ref{fig:osc_exact}. We note that such a situation would always be the case in realistic models, where time dependent backgrounds take very complicated form. Therefore, we should make a strategy to overcome this difficulty: First, we look for an approximated frequency, which has an analytically accessible form and imitates the {\it shape} of the exact one. The fact that the ``area'' in Fig.~\ref{fig:osc_exact} gives the correct particle number supports our strategy.  Therefore, we make approximations for the Hubble parameter as well as the scale factor from the exact expression \eqref{Hosc}, and use them to make analytic estimation. For the scale factor, we numerically integrated \eqref{Hosc} and found that $a(t)=(1+Mt/4)^{\frac23}$ is a very good approximation. By using this approximated scale factor, the effective frequency~\eqref{Omegak} can be rewritten as
\begin{equation}
    \Omega_k^2(t) \approx \frac{k^2}{(1+Mt/4)^{\frac43}} + m^2 - \left(\xi-\frac14\right)\frac{4M^2}{(Mt+4)}\sin(Mt), \label{Omegaapprox}
\end{equation}
where we have truncated the sub-leading order terms in late time limit $Mt\gg1$. Although we see some discrepancy between this approximated form and the exact one derived from \eqref{Hosc} as shown in Fig.~\ref{fig:oscR} for small $Mt$, this approximation becomes better and better in later time $Mt\gg1$. In the following, we discuss the particle production with this approximated formula and will show that an analytic estimate based on \eqref{Omegaapprox} is in agreement with the particle number derived from numerical integration with a more complicated form~\eqref{Hosc}.

At late time oscillation $Mt \gg 1$, where the effective frequency \eqref{Omegaapprox} is a good approximation (Fig.~\ref{fig:oscR}), we can further approximate \eqref{Omegaapprox} as
\begin{align}
    \Omega_k^2(t) \approx \frac{k^2}{(Mt/4)^{\frac43}} + m^2 - \left(\xi-\frac14\right)\frac{4M}{t}\sin(Mt) \label{Omegaapprox2}
\end{align}
for simplicity. Since we are now considering the tachyonic regime, in other words, the large amplitude case, we can take
\begin{align}
    \Delta \equiv \left(\xi - \frac14\right)^{-1} \frac{k_p^2 + m^2}{4M^2}Mt , \label{smallD}
\end{align}
as a perturbation parameter much smaller than unity, where $k_p \equiv \frac{k}{(Mt/4)^{\frac23}}$ is a physical wave number. Finally, with these approximations, we can analytically evaluate the integration~\eqref{hombuntac} and obtain approximated particle number produced at $n$-th oscillation as
\begin{align}
    n_{k,n} \approx \exp\left[ 4\sqrt{\frac{4\xi-1}{(2n+\frac12)\pi}} \left( 2{\rm E}\left( \frac{\pi}{4} \Bigr| 2\right) - {\rm F}\left( \frac{\pi}{4} \Bigr| 2\right) \Delta_{2n+\frac12} \right) \right], \label{nkR2tach}
\end{align}
where $\Delta_{2n+\frac12}$ is the value of the small parameter \eqref{smallD} at $Mt = (2n+\frac12)\pi$,  ${\rm E}(\varphi | k^2)$ and ${\rm F}(\varphi | k^2)$ the incomplete elliptic integral of the second kind and the first kind, respectively. The derivation of this formula is given in appendix~\ref{appC1}. In Fig.~\ref{fig:nu_osc}, we find a good agreement between our analytic estimate of the particle number~\eqref{nkR2tach} and that evaluated by numerical solutions of the exact mode equation. We have also checked other parameter regions and found that our analytic estimate shows a good agreement with numerical one as long as $Mt \gg 1, |\xi| \gg 1$ and the amplitude of the Ricci scalar is large. Nevertheless, we should note that there is $\mathcal{O}(1)$ disagreement for low momentum as shown in~\ref{fig:nu_osc}, which would be originated from the discrepancy between our approximated effective frequency~\eqref{Omegaapprox2} and the exact one shown in Fig.~\ref{fig:oscR}.

\begin{figure}[htbp]
\centering
\includegraphics[width=.64\textwidth]{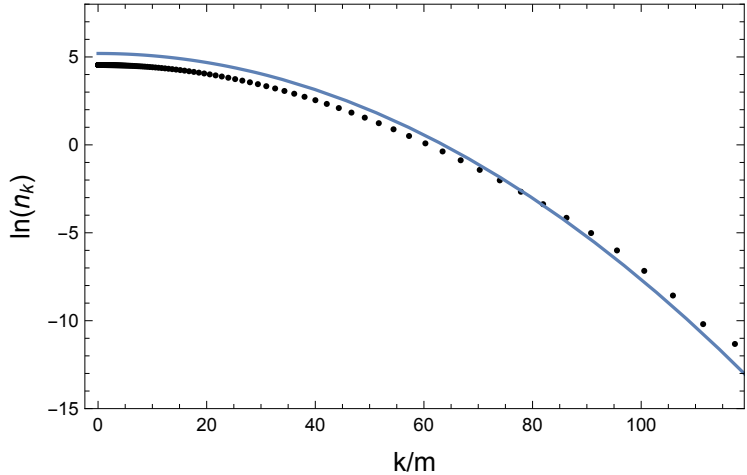}
\caption{\label{fig:nu_osc} Comparison between the numerical result from \eqref{Hosc} and \eqref{Omegak} (black dots) and \eqref{nkR2tach} (blue solid line) at the sixth oscillation $n=5$. In this graph, we take $M = 15.0m, \xi = 10.0$.}
\end{figure}

\begin{figure}[htbp]
\centering
\includegraphics[width=0.64\textwidth]{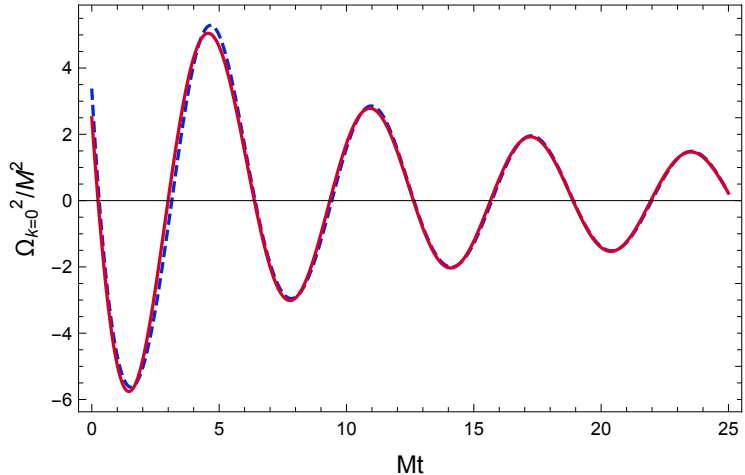}
\caption{\label{fig:improvedR} Comparison between the curvature induced terms of the exact effective frequency derived from the Hubble parameter~\eqref{Hosc} (blue dashed line) and the improved approximation~\eqref{impR} (red solid line). In this graph, we take $k=0, m=0, \xi = 10.0$. The improved one~\eqref{impR} well approximates the exact one even at the first oscillation.}
\end{figure}

\begin{figure}[htbp]
\centering
\includegraphics[width=0.64\textwidth]{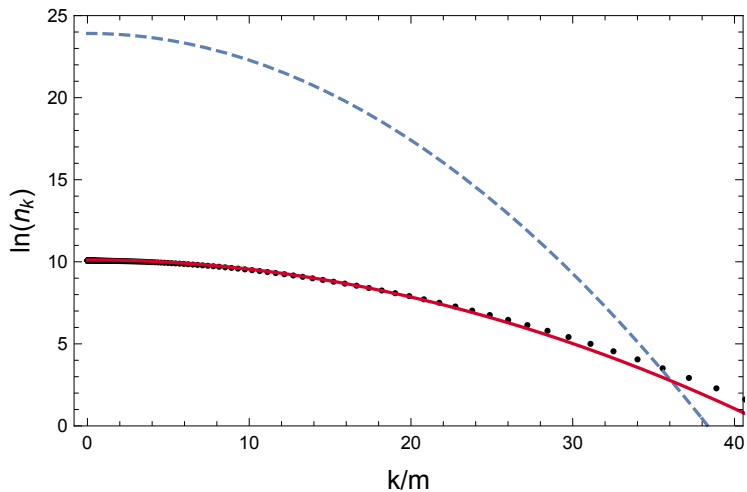}
\caption{\label{fig:compare_osc} Comparison between the first approximation~\eqref{nkR2tach} (blue dashed line) and the improved one~\eqref{analytic osc} (red solid line) at the first oscillation $n=0$. Black dots denote the numerical result from \eqref{Hosc} and \eqref{Omegak}. In this graph, we take $M = 15.0m, \xi = 10.0$. The formula~\eqref{nkR2tach} overestimates the particle number, whereas \eqref{analytic osc} is a good approximation even at the first oscillation.}
\end{figure}

At early time oscillation, especially the first one, the particle number~\eqref{nkR2tach} is not a good approximation as shown by the blue dashed line in Fig.~\ref{fig:compare_osc}. This is simply because the approximated effective frequency~\eqref{Omegaapprox} no longer well-approximates the exact one~\eqref{Omegak} (Fig.~\ref{fig:oscR}). In such a situation, we have to improve the approximation of the effective frequency, which can be achieved by including a sub-leading contribution in~\eqref{Omegaapprox}:
\begin{equation}
    \Omega_k^2 \approx \frac{k^2}{(1+Mt/4)^{4/3}}+m^2-\frac{(4\xi-1)M^2}{Mt+4}\sin (Mt)+\frac{4\xi M^2}{(Mt+4)^2} .\label{impR}
\end{equation}
This improved approximation shows a better agreement with the exact one as depicted in Fig.~\ref{fig:improvedR}. We can make an analytic estimate in the same way as done in previous case, and the analytic estimate agrees with the numerical result even at the first oscillation as shown in Fig.~\ref{fig:compare_osc}. In Appendix \ref{appC2}, we give the detailed derivation of analytic estimation used in Fig.~\ref{fig:compare_osc}. Thus, we have explicitly shown that we are able to find an analytic estimate even if the background is very complicated, and accuracy of the estimate is directly related to that of the effective frequency as we naively expected. We note that the reason why we did not take \eqref{impR} from the beginning is that we would like to explicitly show how the accuracy of the approximation of the effective frequency affects the estimate.

This particular example gives an important lesson: For estimation of the produced particle number, we have first used a roughly approximated form of the effective frequency~\eqref{Omegaapprox}, which has a shape similar to the one given by a (more complicated) exact expression~\eqref{Hosc}. Although the detailed functional forms of the exact one and the approximation are different, our estimation is in good agreement with numerical results. This means that as long as the effective frequency behaves similarly, one may use approximated expressions. Actually, the particle number~\eqref{hombuntac} depends only on the area of the tachyonic region of the effective frequency. This seems very reasonable since $\phi$ experiences the similar backgrounds. However, the agreement between our approximation and the numerical results is not so trivial since the detailed Stokes line structure could be different. In general, a complicated background such as~\eqref{Hosc} may give multiple turning points, which do not show up in our approximated one. What we have shown in this example implies that it is important for good analytic estimation to capture dominant time dependence, namely the dominant Stokes lines and turning points when we approximate the effective frequency. This lesson is crucial for practical purposes since in various models, the time dependence can be very complicated and then it is almost impossible to analyze the Stokes line structure exactly. Nevertheless, we should note that the accuracy of the approximation of the background is very important since the particle density is basically given by an exponential of the phase integral, namely the number density has exponential sensitivity to the effective frequency. This point is clear from Fig.~\ref{fig:compare_osc}.

\section{Smooth transition model: consideration of poles}\label{transition}
In this section, we discuss particle production associated with the transition of the expansion law of the Universe. When the dominant energy component changes e.g.~from matter to radiation, behavior of the scale factor changes and such a transition is known to induce gravitational particle production~\cite{Parker:1969au,Zeldovich:1971mw}. In particular, for such a particle production, the transition time scale determines the efficiency of 
the particle production since the production rate is typically proportional to the factor of the form $e^{-\omega_k\Delta t}$ where $\omega_k$ and $\Delta t$ denote the effective frequency and the transition time scale, respectively~\cite{Chung:1998zb,Hashiba:2018iff,Hashiba:2019mzm}.

We would like to reconsider such a particle production, particularly focusing on the Stokes phenomenon. In general, however, models of smooth transition of the scale factor lead to very complicated effective frequencies, which make analytical study rather difficult. For example, we would model smooth transitions by introducing exponential type functions that smoothly connect different expansion laws in a scale factor, which shows technical difficulties due to the existence of poles as we will explain below. Therefore, it would be more instructive to discuss a simple example having the similar technical difficulty, and to discuss how to treat it within the WKB/phase integral method. Here, we consider a well known toy model with the following effective frequency, which was first investigated in~\cite{Bernard:1977pq}:
\begin{align}
    \omega_k^2 = k^2 + \frac{m^2}{1+e^{-{\frac{t}{\Delta t}}}}. \label{tanh_model}
\end{align}
In this model, the effective frequency changes from $\omega_i^2=k^2$ to $\omega_f^2=k^2+m^2$ mostly around $t=0$ within the time interval $\Delta t$. This toy model would give us an insight on the transition of the scale factor between different era. The time interval $\Delta t$ would correspond to the transition time scale, which may be very short, and then the effective mass coming e.g.~from a curvature term can change suddenly. In this toy model, even if the transition is very sudden, we are able to discuss the particle production rate since the exact solution of the mode equation is known, which yields
\begin{align}
    \beta_k &= \sqrt\frac{\omega_f}{\omega_i} \frac{\Gamma[1-(2{\rm i} \omega_i \Delta t)] \Gamma(2{\rm i} \omega_f \Delta t)}{\Gamma({\rm i} \omega_- \Delta t) \Gamma[1+({\rm i} \omega_- \Delta t)]}, \\
    n_k &= |\beta_k|^2 = \frac{\sinh^2(\pi\omega_- \Delta t)}{\sinh(2\pi\omega_i \Delta t) \sinh(2\pi\omega_f \Delta t)}, \label{tanhexact}
\end{align}
where $\omega_- = \omega_f - \omega_i$, and $\Gamma(z)$ is the complete gamma function. We note that the effective frequency~\eqref{tanh_model} is realized as that of a conformally coupled massive scalar field in a expanding universe where the metric is given by $ds^2=a^2(\eta)(-d\eta^2+d{\bf x}^2)$ and $a^2(\eta)=\frac{1}{1+e^{-\eta/\Delta \eta}}$.\footnote{This interpretation would not be realistic, since in such a spacetime, $\eta\to-\infty$ is an initial singularity, and we are not sure if a scalar particle is well-defined. However, our analysis can be applied to the case where $\omega_{k}^2=k^2+M^2+\Delta m^2(A+B\tanh \frac{\eta}{\Delta\eta})$ by a simple replacement of parameters. This corresponds to the scale factor $a^2(\eta)=(A+B\tanh \frac{\eta}{\Delta\eta})$. There, the initial singularity is absent as long as $A\pm B>0$.}

With possible generalization to expanding Universe in mind, let us calculate the Bogoliubov coefficient $\beta_k$ with the WKB method, and check whether we can capture the particle production caused by the transition. In particular, we first discuss a simple approximated formula, which is often used and gives a certain limit of the exact one~\eqref{tanhexact}. However, we will find an obvious contradiction. We also show how to improve the formula by careful analysis of the Stokes lines.

For the effective frequency~\eqref{tanh_model}, turning points appear at
\begin{align}
    t_c =\left[{\rm i}(2n+1)\pi -\ln\left(\frac{k^2+m^2}{k^2}\right)\right]\Delta t,
\end{align}
where $n$ is an arbitrary integer. Besides the turning points, the effective frequency~\eqref{tanh_model} has an infinite tower of poles at $t = {\rm i}(2n+1)\pi\Delta t$. We should emphasize that the appearance of infinite number of poles and turning points lead to infinite number of Stokes lines as shown below, which make analysis quite difficult in general.

\begin{figure}[htbp]
\centering
\includegraphics[width=.60\textwidth]{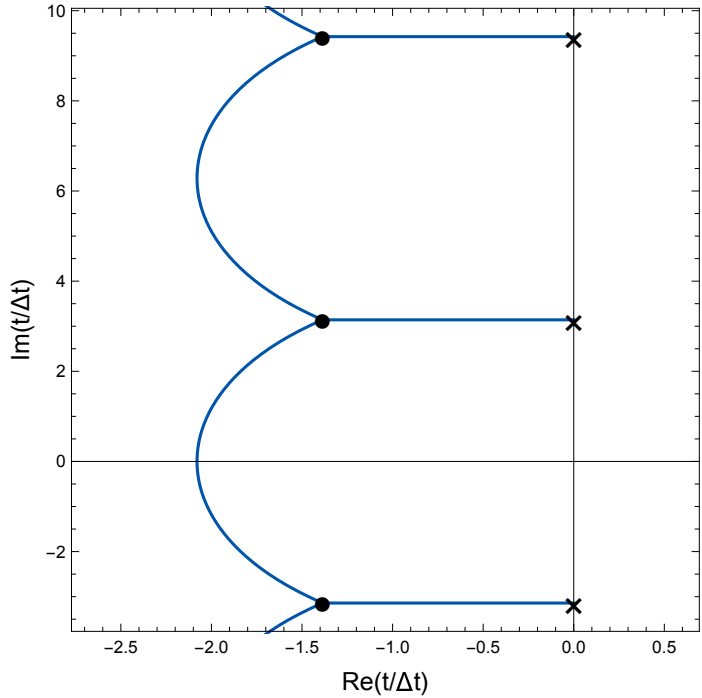}
\caption{\label{fig:st_sltower} Structure of turning points (black dots), poles (black cross marks) and Stokes lines (blue lines) in the model with \eqref{tanh_model}. In this plot, we take $\omega_f = 2\omega_i$. The only Stokes line crossing real time axis is the one connecting the pair of the turning points closest to real $t$-axis.}
\end{figure}

\begin{figure}[htbp]
\centering
\includegraphics[width=.72\textwidth]{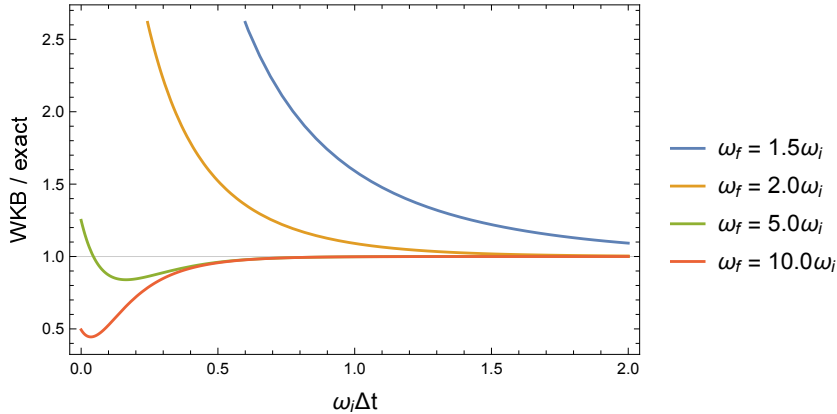}
\caption{\label{fig:WKBfailure} Ratio of the WKB solution~\eqref{tanhWKB} to the exact one~\eqref{tanhexact} with various $\omega_f$. The horizontal line denotes unity, which means two solutions agree with each other. At the adiabatic limit $\Delta t \to \infty$, the WKB solution agrees with the exact one.}
\end{figure}

How can we evaluate the particle production in the presence of {\it infinite} number of turning points? Let us take the following first approximation: Although an infinite number of turning points and poles appear, the most relevant one for particle production is a pair of turning points closest to the real time axis~\cite{Dabrowski:2014ica}. The turning points, poles and Stokes lines in our toy model~\eqref{tanh_model} are depicted in Fig.~\ref{fig:st_sltower}.
Neglecting the Stokes lines associated with poles and turning points with $n\neq 0,-1$, the situation is the same as the model in Appendix~\ref{appA2}.\footnote{One would need to modify the frequency so that Stokes segment disappears as explained in Appendix~\ref{appA2}. In the following, we simply use the result in Appendix~\ref{appA2}.} As explained in Appendix~\ref{appA2}, integration of the effective frequency $\omega_k(t)$ along the Stokes line gives the Bogoliubov coefficient, and accordingly produced particle number density, which in this case is given by
\begin{align}
    n_k = \exp\left(2{\rm i} \int_{t_c}^{\bar{t}_c}\omega_k dt\right) = e^{-4\pi\omega_i \Delta t}. \label{tanhWKB}
\end{align}
We find that this corresponds to \eqref{tanhexact} in the adiabatic limit $\Delta t \to \infty$ while $\omega_i\Delta t$ fixed. If the transition is abrupt $\omega_i \Delta t \lesssim 1$, this WKB solution overestimates or underestimates the particle number for $\omega_f \lesssim 5\omega_i$ or $\omega_f \gtrsim 5\omega_i$, respectively (see Fig.~\ref{fig:WKBfailure}). Nevertheless, this result is consistent with high momentum behavior of the exact solution. The reason can be understood in terms of Stokes lines as follows: In the limit $\Delta t\to \infty$, the Stokes line emanating from the turning points nearest to the real axis seems like a single line and the contributions from other turning points and poles are negligible because the interval is proportional to $\Delta t$. Physical interpretation is that high momentum modes do not feel any event caused by the transition of the mass, and its production rate is independent of mass parameter $m$.

Although \eqref{tanhWKB} gives an approximate formula consistent with the exact solution in  the adiabatic limit, the approximated formula given by the integration along the Stokes line connecting turning points gives non-vanishing value even in the static limit $m \to 0$, which is obvious contradiction. In this sense, the formula~\eqref{tanhWKB} needs to be improved.

\begin{figure}[htbp]
\begin{minipage}{0.48\hsize}
\begin{center}
\includegraphics[width=\textwidth]{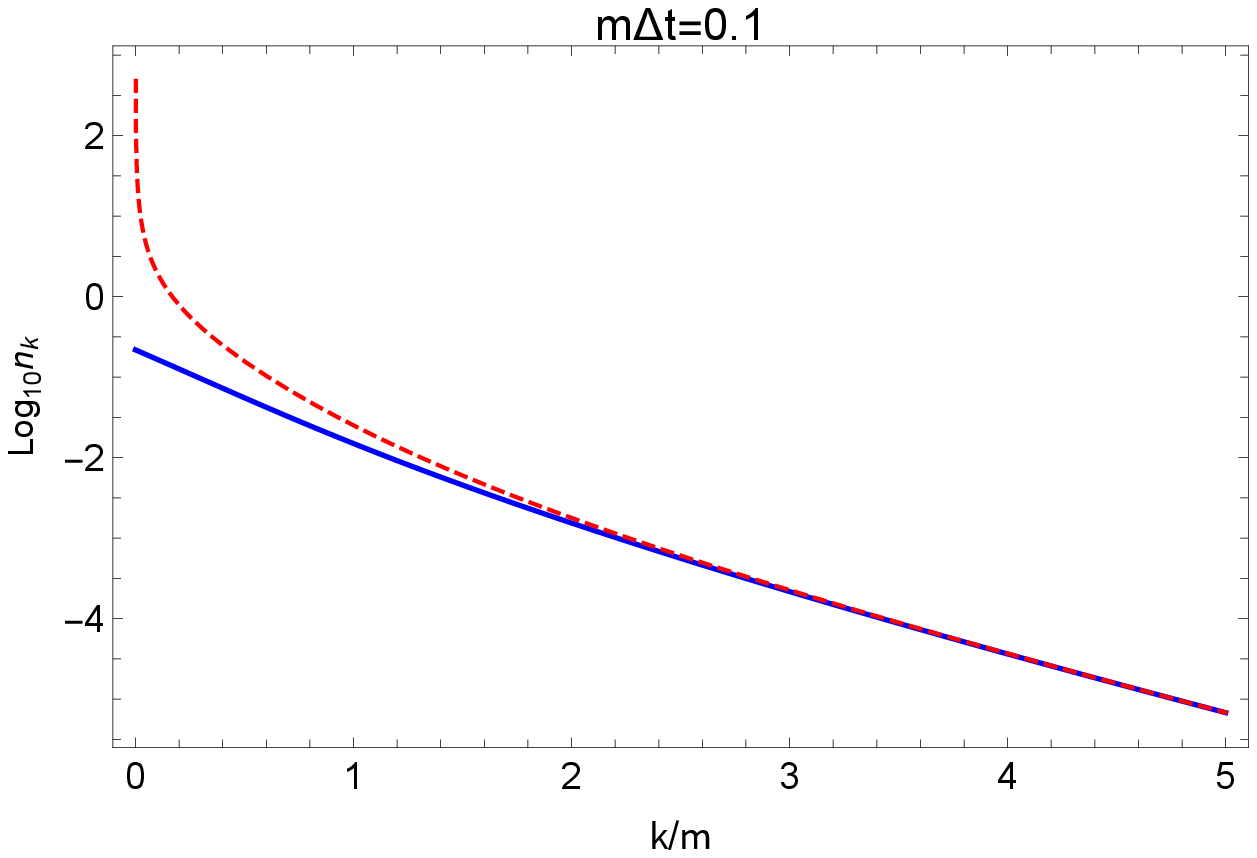}
\end{center}
\end{minipage}
\begin{minipage}{0.48\hsize}
\begin{center}
\includegraphics[width=\textwidth]{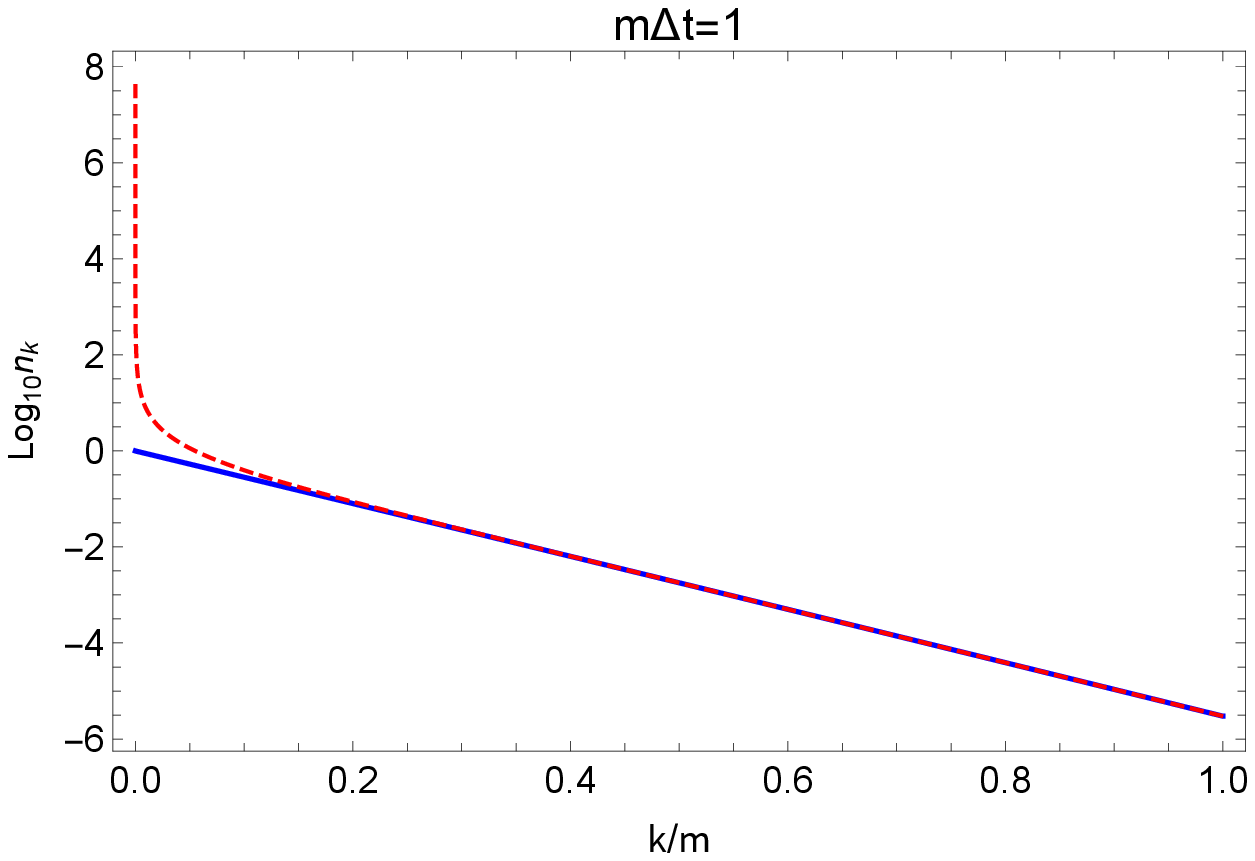}
\end{center}
\end{minipage}
\caption{\label{fig:delta}Comparison of particle number density derived by WKB method~\eqref{improved} and the exact one~\eqref{tanhexact} for $m\Delta t=0.1$ (left) or $m\Delta t=1$(right). The blue solid line is our improved WKB result~\eqref{improved} and the red dashed line is the exact one~\eqref{tanhexact}.}
\end{figure}

How can we improve the formula such that the obvious contradiction is absent while keeping the consistent adiabatic limit? The missing contribution here is that from poles neighboring turning points. Actually, taking more careful treatment of the pole contribution to the Stokes constants\footnote{For the definition of the Stokes constants, see Appendix~\ref{appA}.}, we are able to obtain an improved formula
\begin{equation}
    n_k = \left(1-e^{-2\pi\omega_-\Delta t}\right)^2 e^{-4\pi\omega_i \Delta t}. \label{improved}
\end{equation}
Since the derivation of the improved formula is technical, we discuss the details of its derivation in Appendix~\ref{appB}. The improved formula~\eqref{improved} shows at least a consistent behavior in the limit $m\to0$, at which the produced particle number vanishes as expected since nothing happens in the limit. In the adiabatic limit $k\Delta t>1$, we find expected adiabatic behavior
\begin{equation}
    n_k \sim  e^{-4\pi \omega_i\Delta t}.
\end{equation}
Therefore, the formula~\eqref{improved} has properties that are necessary for consistency. We show the behavior of the improved particle density~\eqref{improved} and exact one~\eqref{tanhexact} with $m=1,\ \Delta t=0.1$ and $m=1,\ \Delta t=1$ in Fig.~\ref{fig:delta}. As expected, the high momentum behavior is precisely reproduced. However, for small $\Delta t$ or low momentum modes, the WKB formula underestimates the particle density. This is reasonable because in both cases, the adiabaticity is violated significantly. Nevertheless, for relatively large momentum limit, the improved WKB formula gives a better approximation for the estimation of the produced particle density. 

We note that similar models were discussed in~\cite{Kim:2013jca} where the author used a different method to evaluate the amount of particle production due to the Stokes phenomenon, with focusing on the contribution coming from the poles in complex time plane. In our analysis, we followed a standard method in evaluating the Stokes constant and continuation of mode functions. 

As we have shown, even in this simple model, one needs to carefully take poles into account when considering the Stokes constants and the connection problem in order to find out the correct particle production rate. Nevertheless, we have also shown how to treat a pole contribution and that such an analysis can in principle give a very good approximation for the produced particle number. Even in more involved situation, similar discussion would enable us to have a good analytic estimate.

However, we should mention that at the abrupt transition limit $\Delta t\to 0$, even the improved formula~\eqref{improved} fails to reproduce the exact one~\eqref{tanhexact} as can be seen e.g.~from $m\Delta t=0.1$ case in Fig.~\ref{fig:delta}. This is because we have taken into account only the nearest two turning points and a pole (see Appendix~\ref{appB}). In the sudden transition limit $\Delta t\to 0$, we can no longer neglect other turning points and poles since the distance between them is scaled by $m\Delta t$. If one would like to consider such a limit, the evaluation with Stokes lines seem not appropriate and different methods would be necessary.

\section{Summary and conclusion}\label{conclusion}
In this paper, we have demonstrated how we can analytically evaluate particle production in time-dependent backgrounds by focusing on the Stokes phenomenon. The Stokes phenomenon analysis clarifies how we should connect solutions of a second order differential equation in regions that have different asymptotic forms of solutions. Since the origin of particle production is nothing but the difference between the asymptotic behavior of mode functions at an early and late time, particle production can be systematically analyzed from the Stokes phenomenon viewpoint. More concretely, in most cases, the amount of produced particle can be evaluated by the phase integral along a Stokes line connecting a pair of turning points, at which the effective frequency vanishes, as shown e.g.~in Eqs.~\eqref{hombuntac} (in tachyonic case) and \eqref{tanhWKB} (in non-tachyonic case).

Exact forms of effective frequency in most realistic models such as that in Sec.~\ref{oscillation} are too complicated to be integrated analytically. Furthermore, even if we find turning points and Stokes lines, the number of them can be infinite as the model in Sec.~\ref{transition}. In such cases, complete analysis of the Stokes phenomenon seems impossible or impractical. Therefore, for practical purposes, we need to know whether we can use approximations in order to obtain analytic estimation for such cases. 

In this paper, we have addressed this issue and discussed possible approximations for practical uses of the Stokes phenomenon analysis. In Sec.~\ref{oscillation}, we have discussed how the approximation of the effective frequency affects an estimate of particle production rate. We have found that our simplified effective frequency can give us an estimate of satisfactory accuracy. We have also shown that the improvement of an approximated frequency is directly related to accuracy of an analytic estimate. In Sec.~\ref{transition}, we have discussed how to treat an infinite number of the turning points as well as poles. Despite the infinite number of them, we have found that the Stokes line connecting the turning points nearest to the real time axis gives a dominant contribution to the particle production. As shown in Sec.~\ref{transition}, taking into account only this Stokes line gives us an estimate consistent with adiabatic limit. Furthermore, we have also found that the estimate can be improved by taking into account the contribution from poles. Although we have only taken into account one pole contribution among infinite number of them, we have found that such an improvement makes much better the agreement between the analytic estimate and the exact solution.

Our observations mean that, for analytic estimates, we need neither exact functional forms of background nor analysis of all the Stokes lines and poles. An approximated effective frequency and the contribution from the Stokes line and pole that are closest to the real time axis are enough to evaluate the amount of particle production. Our result would be useful to analyze particle production even in more complicated situation including not only gravitational particle production but also general non-perturbative production such as the Schwinger effect and the Hawking radiation.

In conclusion, produced particle number can be well estimated in many cases by the following recipe: Approximate the effective frequency so that it can be integrated analytically. An approximation of the effective frequency would work as long as its shape and the structure of Stokes lines are similar to the exact one. Find turning points and Stokes lines of the approximated effective frequency in the complex time plane. Then, perform the phase integral of the approximated effective frequency along the Stokes segments and we will obtain an estimate of produced particle number. If there are infinite numbers of Stokes lines, take into account the one crossing the real time axis, as first approximation. In the presence of poles, take the one closest to a real axis into account for the next order approximation as we have done in appendix~\ref{appB}.

We admit that the WKB/phase integral method fails at non-adiabatic limits such as an abrupt transition (e.g.~$m\Delta t \to 0$ limit in Figs.~\ref{fig:WKBfailure} and \ref{fig:delta}) since our analysis is based on the WKB solution, namely, the adiabatic approximation. Although even in such a case we can improve our estimation by taking into account the contribution from poles and other Stokes lines, there might be more appropriate mathematical methods for such situations. We leave it for our future study.

Finally, we would like to mention other possible future directions. The time evolution of the Bogoliubov coefficients e.g.~during transition, can be described by an optimal approximation shown by Dingle and derived by Berry~\cite{Barry:1989zz} (for review, see e.g.~\cite{Dabrowski:2016tsx,Li:2019ves}). With our approximation methods, such a time dependent formula allows us to discuss, for instance, backreaction from particle production to the background. It would also be interesting to see the relation to the world line instanton method~\cite{Dunne:2005sx}. We expect that such a different formulation gives us a complementary insight on the particle production in cosmological models. We will address these possibilities in future.

\section*{Acknowledgement}
We would like to thank Jun'ichi Yokoyama for discussion. SH is supported by JSPS KAKENHI, Grant-in-Aid for JSPS Fellows 20J10176 and the Advanced Leading Graduate Course for Photon Science (ALPS). YY is supported by JSPS KAKENHI, Grant-in-Aid for JSPS Fellows JP19J00494.

\appendix

\section{Review of WKB/phase integral method and Stokes phenomenon}\label{appA}
In this appendix, we give a brief review of the phase integral method with particular focus on the Stokes phenomenon, which is crucial for understanding non-perturbative particle production. (For a more comprehensive review, see e.g.~\cite{Froman1}.) We also note that recently the exact WKB analysis is applied to particle production context~\cite{Enomoto:2020xlf,Taya:2020dco}. The exact WKB analysis is mathematically more rigorous, but for our practical purposes, the usual WKB method, which we will use in this work seems sufficiently useful. The phase integral method can be thought of generalization of the WKB method. To discuss the particle production in time-dependent backgrounds including curved spacetime as well as 
an oscillating scalar field, we need to solve the mode equation of a target particle, but, in general, finding an exact solution is quite difficult or impossible. Let us consider the following second order differential equation,
\begin{equation}
    \ddot{\psi}(t)+\omega^2(t)\psi(t)=0,\label{modeeq}
\end{equation}
where $\omega(t)$ corresponds to a time dependent frequency. We have the following formal solutions
\begin{equation}
\psi(t)=\frac{1}{\sqrt{q(t)}}\exp\left(\pm {\rm i}\int^tdt'q(t')\right). \label{pi}
\end{equation}
Here, $q(t)$ is an unspecified function satisfying
\begin{equation}
    q^{-3/2}\frac{d^2q^{-1/2}}{dt^2}+\frac{\omega^2}{q^2}-1=0.\label{exact}
\end{equation}
If such $q(t)$ is found, it means that we find an exact solution, but as mentioned above, it is not always the case. Instead, let us consider an approximate solution
\begin{align}
    \psi(t) \approx \frac{1}{\sqrt{Q(t)}}\exp\left(\pm {\rm i}\int^tdt'Q(t')\right), \label{WKBsoln}
\end{align}
where $Q(t)$ is a chosen function to obtain an approximate solution. In order to quantify the error of the approximation is, we introduce the following quantity
\begin{align}
    \varepsilon_0(t)\equiv& Q^{-3/2}\frac{d^2Q^{-1/2}}{dt^2}+\frac{\omega^2}{Q^2}-1\nonumber\\
    =&\frac{1}{16Q^6}\left[5\left(\frac{dQ^2}{dt}\right)^2-4Q^2\frac{d^2Q^2}{dt^2}\right]
    +\frac{\omega^2-Q^2}{Q^2}.\label{eps}
\end{align}
The value of $|\varepsilon_0|$ shows the error of the approximate solution~\eqref{WKBsoln}. In principle, one can choose $Q(t)$ to be whatever function as long as the approximation is valid, namely, $|\varepsilon_0|\ll1$. The (zeroth order) WKB solution corresponds to the special choice $Q(t)=\omega(t)$. Sometimes, the freedom of the choice of $Q(t)$ is useful to have a better approximation than the WKB one. If the time derivatives of $\omega(t)$ are sufficiently small, one may think of $\varepsilon_0$ and its derivatives as small quantities for the WKB solution. Starting with the WKB solution $Q(t)=\omega(t)$ as the lowest order, we can recursively solve \eqref{eps} with respect to $Q(t)$. Then, one would find $Q(t)$ including higher-order time derivative terms in $\varepsilon_0$. The role of higher-order terms are discussed e.g.~in~\cite{Froman1,Kim:2007pm}. However, for simplicity, we do not discuss the improvement with higher-order contributions here. We note that one would be able to construct a mode function including infinite series of higher order terms, which naively seems an exact solution. However, such a series cannot give an exact solution since it is an asymptotic series. In this regard, one needs to take into account the Stokes phenomenon, which is explained below.

Once we find a good approximated solution, we expect that the solution can be continued along whole (real) time axis. However, this is not true in general. If we extend the time $t$ to be complex, generally there are points $Q(t)=0$, which are called {\it turning points}. At turning points, the error \eqref{eps} becomes large (or even diverges) and the approximate solution \eqref{WKBsoln} is invalid there. Such behavior may also happen if $Q(t)$ has poles in complex $t$ plane. More importantly, from turning points, so-called Stokes and anti-Stokes lines emanate, which go to infinity or end up at the poles or other turning points. The Stokes and anti-Stokes lines are defined as follows:
\begin{align}
\text{Stokes line: }&Q(t)dt=\text{pure imaginary},\\
\text{anti-Stokes line: }&Q(t)dt=\text{real}.
\end{align}
Note that every Stokes and anti-Stokes line from the same turning point never intersects with each other. Stokes lines are particularly important for discussion of the particle production. On the Stokes line, $\exp(\pm {\rm i}\int Q(t)dt)$ increases or decreases significantly. It is well known that the WKB/phase integral solution is given as asymptotic series and defined only locally, as mentioned above. The Stokes line can be thought of the boundary of the regions, each of which has locally defined solutions.\footnote{This is more precisely discussed in the context of the exact WKB analysis. In the exact WKB analysis, one considers the Borel resummation of the asymptotic series, which gives the exact solution from the asymptotic series. However, on Stokes lines, the solutions are not Borel summable, and in this sense, a Stokes line is actually the boundary of locally defined solutions.} Therefore, in order to know the global behavior of solutions, one has to consider the connection of the solutions defined in each region separated by Stokes lines. We can locally define ``positive'' and ``negative'' frequency modes, and the connection of the solutions leads to mixing of the positive and negative frequency modes, which is the {\it Stokes phenomenon}. Even if there exists no negative frequency mode in one region, the Stokes phenomenon may give rise to the negative frequency mode in another region. Physical interpretation of this is nothing but particle production. It is important to note that even if turning points are not on the real time axis, Stokes lines emanating from them may cross the real time axis and lead to the change of the mode function behavior in real time as we will discuss in detail hereafter.

\begin{figure}[htbp]
\centering
\includegraphics[width=.45\textwidth]{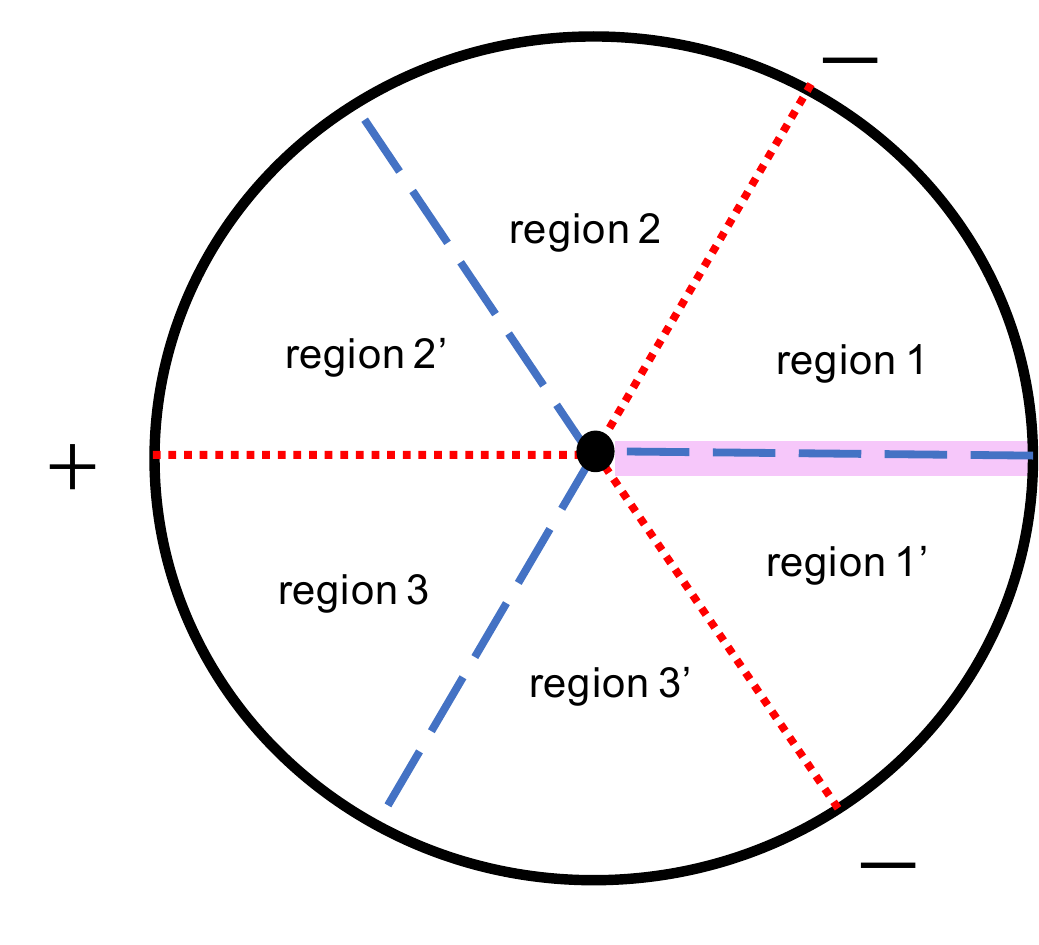}
\caption{\label{fig:ex} Stokes and anti-Stokes lines around a turning point. The center dot is a turning point, red dotted  (blue dashed) lines represent (anti-)Stokes lines. The shaded line corresponds to a branch cut. The plus and minus signs mean that $+{\rm i}\int_{t_c}^t Q(t')dt'$ increases ($+$) or decreases ($-$) along the Stokes line. }
\end{figure}

We briefly summarize the Stokes phenomenon as follows. First, we assume that the two basis solutions are given by
\begin{equation}
    f_{\pm}(t) = \frac{1}{\sqrt{Q(t)}}\exp\left(\pm{\rm i}\int_{t_0}^tQ(t') dt'\right), \label{fpm}
\end{equation}
where $t_0$ is some arbitrary point in complex $t$-plane. Suppose there is a simple turning point $t_c$ around which $Q^2(t)$ can be expanded as
\begin{equation}
    Q^2(t) = A(t-t_c)+{\cal O}\left( (t-t_c)^2 \right),
\end{equation}
where $A$ is a non-vanishing constant. Let us take the lower limit of phase integral $t_0$ to be $t_c$, and the exponent in~\eqref{fpm} can be approximated as
\begin{equation}
    \pm{\rm i}\int_{t_c}^tQ(t')dt' \sim \pm{\rm i}\sqrt{A}(t-t_c)^{\frac32}
\end{equation}
around the turning point. This approximation tells us that there are three Stokes and three anti-Stokes lines respectively. More specifically, parametrizing $t$ as $t-t_c=r e^{{\rm i}\theta}$ where $r$ and $\theta$ are real ($0<r, 0\leq\theta<2\pi$), the Stokes lines correspond to $\theta=\frac{\pi}{3},\pi,\frac{5\pi}{3}$, and the anti-Stokes lines $\theta=0,\frac{2\pi}{3},\frac{4\pi}{3}$. Note that this approximation is valid only around $t=t_c$ and the lines are in general not straight. Schematically, the structure of Stokes and anti-Stokes lines around a turning point are depicted in Fig.~\ref{fig:ex}.
Stokes lines separate a region surrounding a turning point into three parts in general. As we show in Fig.~\ref{fig:ex}, each Stokes line is characterized by whether $+{\rm i}\int_{t_c}^t Q(t')dt'$ increases or decreases on the line, which is specified by plus or minus signs in Fig.~\ref{fig:ex}. These signs are determined by our choice of the phase of $Q(t)$ around the turning points. For instance, along the minus Stokes line, $f_+$ decreases exponentially whereas $f_-$ increases. In this case, we call $f_-$ to be dominant, and $f_+$ subdominant. For the plus Stokes line, the dominant and subdominant components are opposite. When crossing an anti-Stokes line, the dominant and subdominant component change. For example, in Fig.~\ref{fig:ex}, $f_-$ is dominant in the region 2 and becomes subdominant in region $2'$. One may also think that $f_+$ becomes dominant because it is near a plus Stokes line.\footnote{In this sense, only Stokes lines are important to know which component is dominant as well as to know where Stokes phenomenon takes place. Therefore, one can focus only on Stokes lines. Here, we have shown anti-Stokes lines to clarify their role.} When we consider the continuation of the solution in one region to another separated by a Stokes line, the Stokes phenomenon takes place. We briefly summarize the connection rules as follows:
\begin{enumerate}
    \item When we continue, in a counterclockwise sense, a local solution in one region to other separated by a minus (plus) Stokes line, the basis solutions change as
    \begin{equation}
        f_{\mp}\to f_{\mp}+Sf_{\pm},\qquad f_{\pm}\to f_{\pm},
    \end{equation}
    where the upper (lower) sign is for a minus (plus) Stokes line. Here $S$ is called a Stokes constant, which is determined by consistency conditions. We also note that Stokes constants are assigned to each Stokes line. For instance, in Fig.~\ref{fig:ex}, the connection from region 1 to region 2 leads to $f_-\to f_-+Sf_+$ whereas $f_+\to f_+$. When solutions are continued in clockwise direction, the sign of $S$ becomes opposite $S\to -S$.
    \item When we consider a connection at one Stokes line, we have to change the lower limit of the phase integral in the basis solutions~\eqref{fpm} to be the turning point from which that Stokes line emanates. After changing the lower limit, one can apply the above rule. Suppose we have two turning points $t_a$ and $t_b$. When we consider connection of solutions at a Stokes line that emanates from $t_a$, we take the basis solutions to be $f_{\pm}=\frac{1}{\sqrt{Q}}\exp(\pm {\rm i}\int_{t_a}^{t}Qdt')$. If we next consider the continuation of a solution to the region separated by a different Stokes line emanating from $t_b$, we should change the basis solutions as
    \begin{equation}
       f_{\pm} = e^{\pm K}\exp\left(\pm{\rm i}\int_{t_b}^tdtQ(t)\right) \equiv e^{\pm K}\tilde{f}_\pm,
    \end{equation}
    where $K={\rm i}\int_{t_a}^{t_b} Q(t)dt$ connecting the two turning points. When crossing the Stokes line emanating from $t_b$, the connection rule explained above is applied to $\tilde{f}_\pm$ rather than $f_\pm$.
    \item When we consider the crossing of a branch cut in a counterclockwise sense, the basis solutions change as\footnote{This rule originates from the single-valued-ness of the original differential equation. The branch cut appears due to $Q(t)$, which is introduced to (approximately) solve the mode equation. Such a branch cut, however, does not appear for an exact solution, since $\omega^2(t)$ is a single valued function. Therefore, the basis solutions in different Riemann sheets should be related to each other. This observation leads to~\eqref{cutrule}.}
    \begin{equation}
        f_{\pm}\to -{\rm i}f_{\mp}.\label{cutrule}
    \end{equation}
\end{enumerate}
One of the important consequences of the first and the third rule is the following: Suppose a turning point is well separated from any other turning points and poles, and the corresponding Stokes lines are of the form shown in Fig.~\ref{fig:ex}. Let us consider the connection from region 1 to region $1'$ in the counterclockwise direction. One can easily confirm that the connection is given as
\begin{equation}
    \left(\begin{array}{c}f_+\\ f_-\end{array}\right)\to \left(\begin{array}{cc}1&0\\ S&1\end{array}\right)\left(\begin{array}{cc}1&S'\\ 0&1\end{array}\right)\left(\begin{array}{cc}1&0\\ S''&1\end{array}\right)\left(\begin{array}{c}f_+\\ f_-\end{array}\right),
\end{equation}
where $S$, $S'$ and $S''$ are Stokes constants of each Stokes line. On the other hand, we also know that when we cross the cut, the connection in the clockwise direction\footnote{Note that the connection rule~\eqref{cutrule} is applied for the {\it counter}-clockwise direction. Here we consider the opposite direction and then  the sign of ${\rm i}$ changes.} is given by
\begin{equation}
    \left(\begin{array}{c}f_+\\ f_-\end{array}\right)\to \left(\begin{array}{cc}0&{\rm i}\\ {\rm i}&0\end{array}\right)\left(\begin{array}{c}f_+\\ f_-\end{array}\right).
\end{equation}
Since these two connection rules should be the same, we find a consistent solution to be
\begin{equation}
    S=S'=S''={\rm i}.\label{SC}
\end{equation}
This is one of the examples showing how we fix the Stokes constants from consistency conditions. We should emphasize that this result is very much used when we take the well-separated turning points approximation. Namely, if turning points are well separated, we may simply take a Stokes constant to be ${\rm i}$ for counterclockwise connection (or $-{\rm i}$ for clockwise connection).

\subsection{Examples}\label{appAone}
\subsubsection{Particle production from instability}\label{appA1}
Let us show some examples of connection problems in order to understand how the above rules are applied. As the first example, we consider the following effective frequency
\begin{equation}
    \omega^2(t) = t^2-a^2,
\end{equation}
where $a$ is a real positive constant. This toy model describes a particle which experiences instability during a time interval $t \in (-a,a)$. Although the notion of the particle is not well-defined during instability phase, we regard the amplification of the negative frequency mode as ``particle production''. 

When we discuss Stokes lines associated with the effective frequency $\omega^2(t)$, which is real on the real $t$-axis, there can be Stokes lines connecting two turning points, which we call a {\it Stokes segment}. Actually, in this toy model, there is a Stokes segment connecting $t=-a$ and $t=a$ on the real $t$-axis. As discussed e.g.~in \cite{Taya:2020dco}, such a segment makes it difficult to analyse the connection problem.\footnote{In \cite{Taya:2020dco}, this statement is explained in the context of the exact WKB analysis. We also find the similar (or essentially the same) difficulty.} Therefore, we consider a prescription (see e.g.~\cite{Taya:2020dco}): We introduce an infinitesimal phase to the parameter $a\to ae^{{\rm i}\eta}$ and $\eta\neq0$ is a real (small) parameter. Such a modification leads to deformation of the Stokes line structure, and as a result, the Stokes segment is decomposed into two Stokes lines. Depending on the sign of $\eta$, we have two different types of diagrams as shown in Figs.~\ref{fig:tachyonic1} and \ref{fig:tachyonic2}. As we will see below, the difference of the two choices disappear at the leading order of the semiclassical analysis, as shown in \cite{Taya:2020dco}. In other words, such a difference cannot be removed if the semiclassical/adiabatic approximation is invalid, and we may regard it as the limitation on the WKB/phase integral method.

Let us discuss the connection problems for a small positive and negative $\eta$ separately, which clarifies the above statement. The schematic pictures of Stokes lines for each case are shown in Figs.~\ref{fig:tachyonic1} and~\ref{fig:tachyonic2}, respectively.\footnote{The Stokes lines in these figures are described as straight lines, but actual ones are not straight but curved. They are shown only for an illustrative purpose.} In both cases, we take the initial adiabatic vacuum mode function to be
\begin{equation}
    \psi(t) = \frac{1}{\sqrt{\omega(t)}}\exp\left(-{\rm i}\int_{t_0}^t \omega(t') dt' \right) = (t,t_0),
\end{equation}
where $t_0(\ll-{\rm Re}(ae^{{\rm i}\eta}))$ denotes the initial time, and we have used a shorthand notation $(x,y)\equiv\frac{1}{\sqrt{\omega}}\exp\left({\rm i}\int_x^y \omega dt \right)$ introduced in~\cite{White}. 

\begin{figure}[htbp]
\centering
\includegraphics[width=0.65\textwidth]{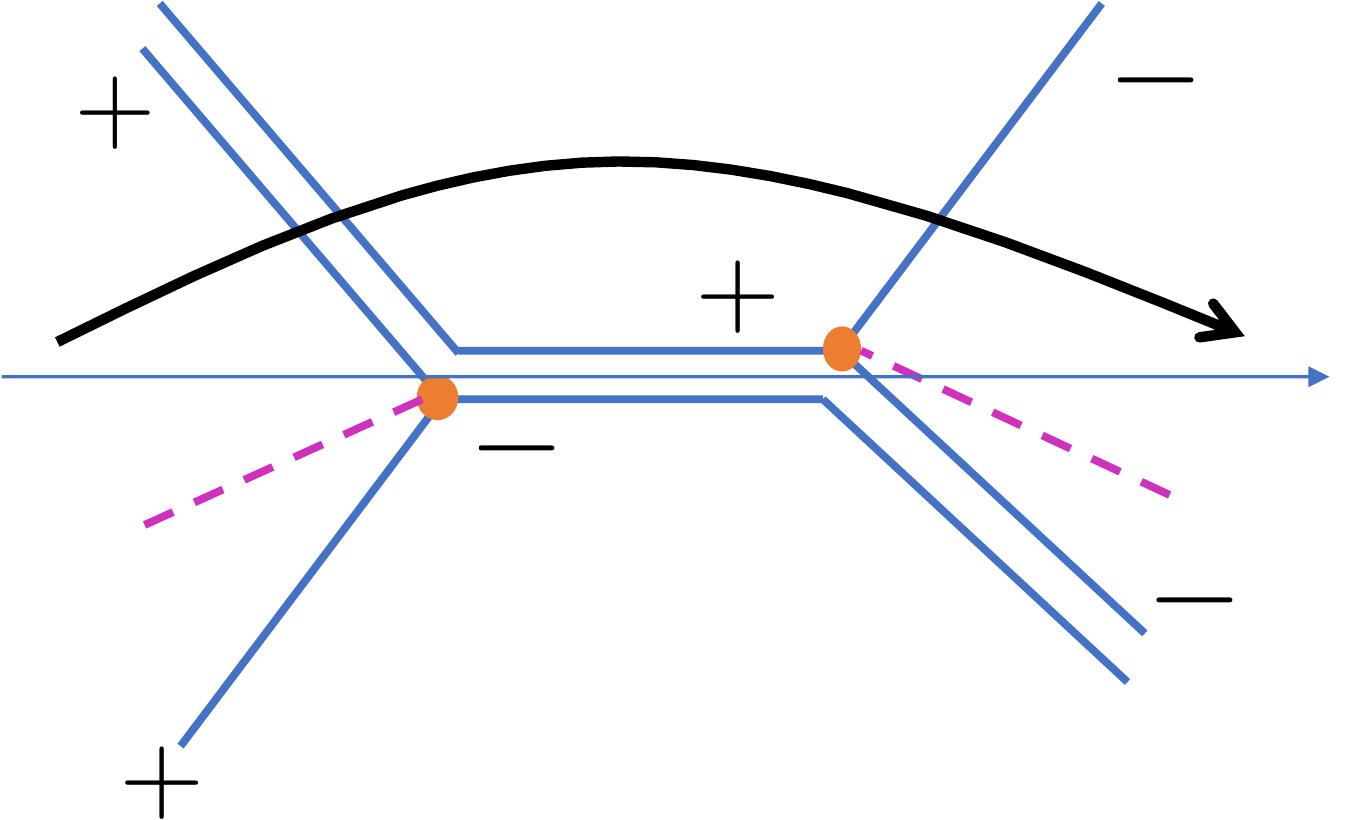}
\caption{\label{fig:tachyonic1} A schematic picture of Stokes line structure for $\omega^2=t^2-a^2e^{2{\rm i}\eta}$ for $\eta>0$. The orange dots are turning points $t=\pm ae^{{\rm i}\eta}$. The solid blue lines are Stokes lines, the purple dashed lines the branch cuts, and horizontal arrow the real $t$-axis. We consider the connection along the black thick arrow. The signs of Stokes lines are also shown.}
\end{figure}
Let us first consider the case with $\eta>0$ ($|\eta|\ll1$) shown in Fig.~\ref{fig:tachyonic1}. In this case, the Stokes line we first cross emanates from a turning point $t=-ae^{{\rm i}\eta}\equiv -t_c$. Then, one needs to change the lower limit of the phase integral according to the second rule, and the mode function is written as
\begin{equation}
    \psi = [-t_c,t_0](t,-t_c),
\end{equation}
where we have introduced a shorthand notation $[a,b]\equiv\exp\left({\rm i}\int_a^{b}\omega dt\right)$. Since the first Stokes line has a positive sign, $(t,-t_c)$ is a subdominant, and therefore, the continuation is done without any change,
\begin{equation}
    \psi\to\psi'=[-t_c,t_0](t,-t_c).
\end{equation}
The second Stokes line emanates from the turning point $t=t_c$, so we need to change the lower limit of the phase integration. Thus, we change the basis functions to be
\begin{equation}
    \psi'=[t_c,t_0](t,t_c).
\end{equation}
Again, we find that the mode function is subdominant, and there is no change:
\begin{equation}
    \psi'\to\psi''=[t_c,t_0](t,t_c).
\end{equation}
Finally, the third Stokes line is a negative one, and the mode function becomes dominant, and therefore, the mode function is continued as
\begin{equation}
    \psi''\to \tilde{\psi}=[t_c,t_0]\left\{(t,t_c)+S(t_c,t)\right\},
\end{equation}
where $S$ is a Stokes constant. Let us separate the factor $[t_c,t_0]$ as $[t_c,-t_c]=e^{\kappa}$ and $e^{{\rm i}\zeta}=[t_0,-t_c]$, where $\kappa$ and $\zeta$ are real. We consider the limit $\eta\to+0$, and such a limit makes $\kappa>0$ since the integral $[t_c,-t_c]$ is along the Stokes line emanating from $t_c$, which has a positive sign. Since the mode function experiences the tachyonic regime, it would be appropriate to change the lower limit of the phase integral to be $t=t_c$. In this sense, the factor $e^{\kappa}$ can be interpreted as the growth factor due to the tachyonic instability. The Stokes constant is not yet determined; however, as we explained below \eqref{SC}, the Stokes constant for a well-separated tuning point is given by $S\sim {\rm i}$, and we take this approximation. Then, the resultant mode function is given by
\begin{equation}
    \tilde{\psi}\sim e^{\kappa+{\rm i}\zeta}\left\{(t,t_c)+{\rm i}(t_c,t)\right\}.
\end{equation}

\begin{figure}[htbp]
\centering
\includegraphics[width=0.65\textwidth]{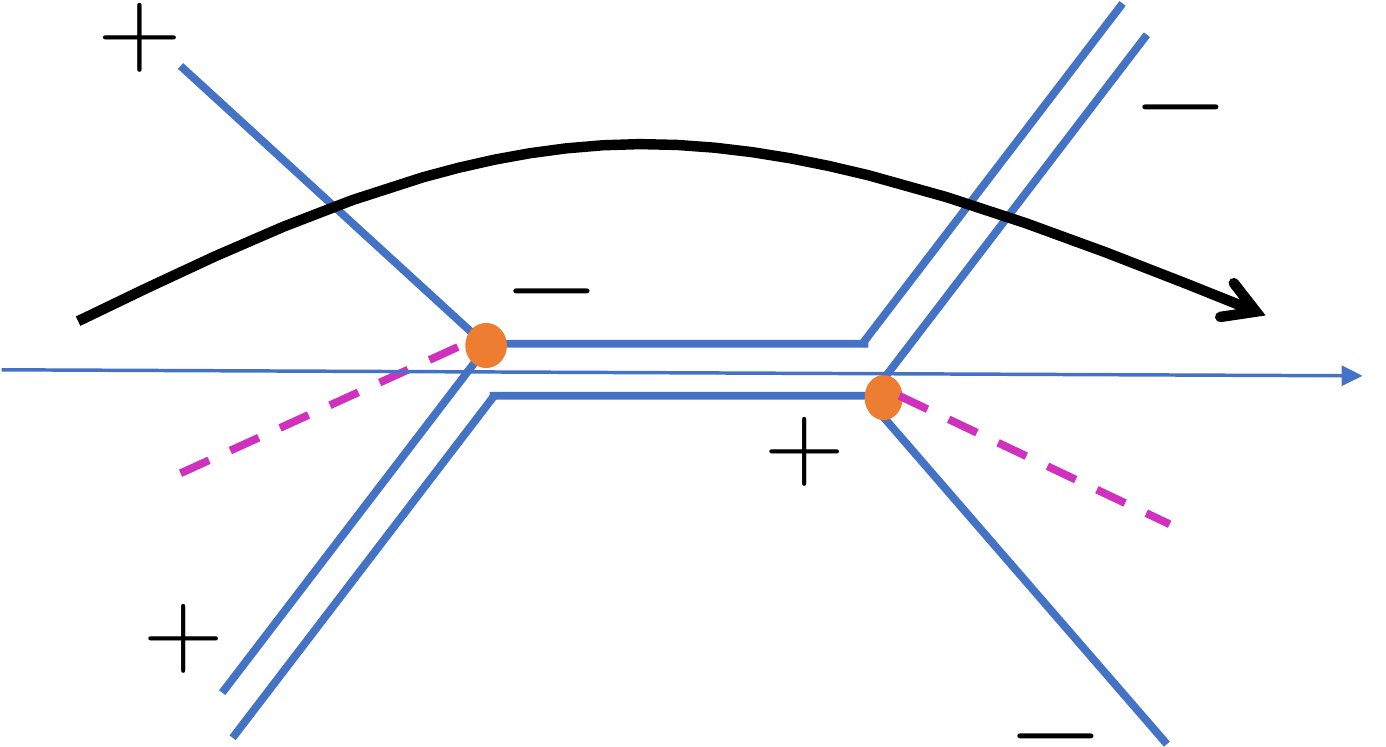}
\caption{\label{fig:tachyonic2} The Stokes line structure for $\omega^2=t^2-a^2e^{2{\rm i}\eta}$ for $\eta<0$.}
\end{figure}
Next, let us turn to the case with $\eta<0$ shown in Fig.~\ref{fig:tachyonic2}. The initial mode function $\psi=[-t_c,t_0](t,-t_c)$ is subdominant for the first Stokes line crossing, and the continued mode function is $\psi'=[-t_c,t_0](t,-t_c)$. The second Stokes line emanates from $t=-t_c$ and we do not change the lower limit of the phase integral, but now the second one has a negative sign. Thus, we find the mode function after the second crossing to be
\begin{equation}
    \psi''=[-t_c,t_0]\left\{(t,-t_c)+S'(-t_c,t)\right\},
\end{equation}
where $S'$ is a Stokes constant, which again we approximate to be $S'\sim {\rm i}$. Finally, for the third Stokes line crossing, we change the lower limit, which leads to 
\begin{equation}
    \psi''=[t_c,t_0]\left\{(t,t_c)+S'([-t_c,t_c])^2(t_c,t)\right\}.
\end{equation}
The continuation of the mode function after the third Stokes line crossing is
\begin{equation}
    \tilde{\tilde{\psi}}=[t_c,t_0]\left\{(t,t_c)+(S''+S'([-t_c,t_c])^2)(t_c,t)\right\},
\end{equation}
where $S''$ is another Stokes constant. After taking the well-separated turning point approximation $S''\sim{\rm i}$ and the limit $\eta\to-0$, we find
\begin{equation}
    \tilde{\tilde{\psi}}\sim e^{\kappa+{\rm i}\zeta}\left\{(t,t_c)+{\rm i}(1+e^{-2\kappa})(t_c,t)\right\}.
\end{equation}
As we mentioned, at the leading order of the semiclassical approximation $\kappa\gg1$, we neglect the small factor $e^{-2\kappa}$ and find
\begin{equation}
    \tilde{\psi}\sim\tilde{\tilde{\psi}},
\end{equation}
and the difference between $\eta\to \pm 0$ disappears.

If we interpret the overall factor of the mode function as the growth of the Bogoliubov coefficients, we find the produced ``particle number'' to be
\begin{equation}
    n=|\beta|^2=e^{2\kappa}=e^{\pi a^2}.
\end{equation}
We should also note that our approximation does not seem to respect the conservation of normalization of the Bogoliubov coefficient $|\alpha|^2-|\beta|^2=1$. This is because we have neglected the subleading order, and if we took more careful analysis of the connection problem, we could find a correct normalization. However, for most of the practical purposes, where particle production is not so drastic, this evaluation is enough for estimation.

It would be also useful to note the following approximation for a practical use: As we saw the above, the particle production associated with the instability was simply given by the integral along the Stokes segment:
\begin{align}
    |\beta| = e^{\kappa} = \exp\left( \int_{-t_c}^{t_c}|\omega| dt \right). \label{eqappA1}
\end{align}
This seems a rather general result when we consider well-separated turning points. Roughly speaking, the particle number density is typically given by the integral along Stokes segments $[t_1,t_2]$ (with an appropriately chosen phase of $\omega$) where $t_{1,2}$ denote turning points connected by a Stokes segment. We actually see that this is true in the next example, which has a different but similar structure of Stokes lines.

\subsubsection{Particle production induced by fast moving background}\label{appA2}
The next example we show is more realistic and well known one. Let us consider the following effective frequency
\begin{equation}
    \omega^2(t) = k^2+v^2t^2,
\end{equation}
where $k$ and $v$ are real constants. This effective frequency describes the model where a scalar field $\chi$ couples to a coherently oscillating scalar field $\phi(t)$ after the end of inflation through an interaction $\lambda^2\phi^2\chi^2$, which is known as the broad resonance regime in preheating. Near $\phi\sim 0$, such an effective mass looks $m^2\sim\lambda^2\dot\phi_*^2t^2$ where $\dot{\phi}_*$ is approximated as a constant, and we identify $v\sim \lambda\dot{\phi}_*$~\cite{Kofman:1997yn,Kofman:2004yc}.

\begin{figure}[htbp]
\centering
\includegraphics[width=0.45\textwidth]{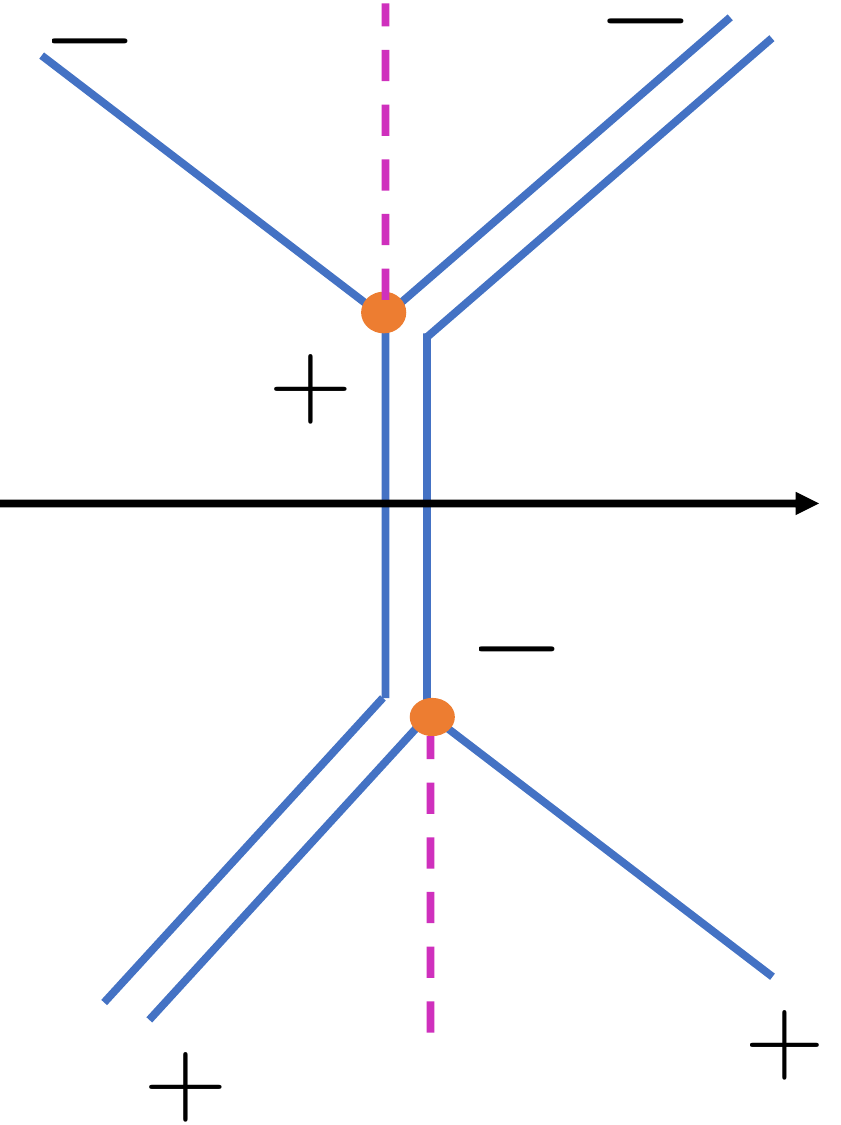}
\caption{\label{fig:nt1} A schematic picture of Stokes line structure of $\omega^2=k^2+v^2t^2$ with $\eta>0$.}
\end{figure}

\begin{figure}[htbp]
\centering
\includegraphics[width=0.45\textwidth]{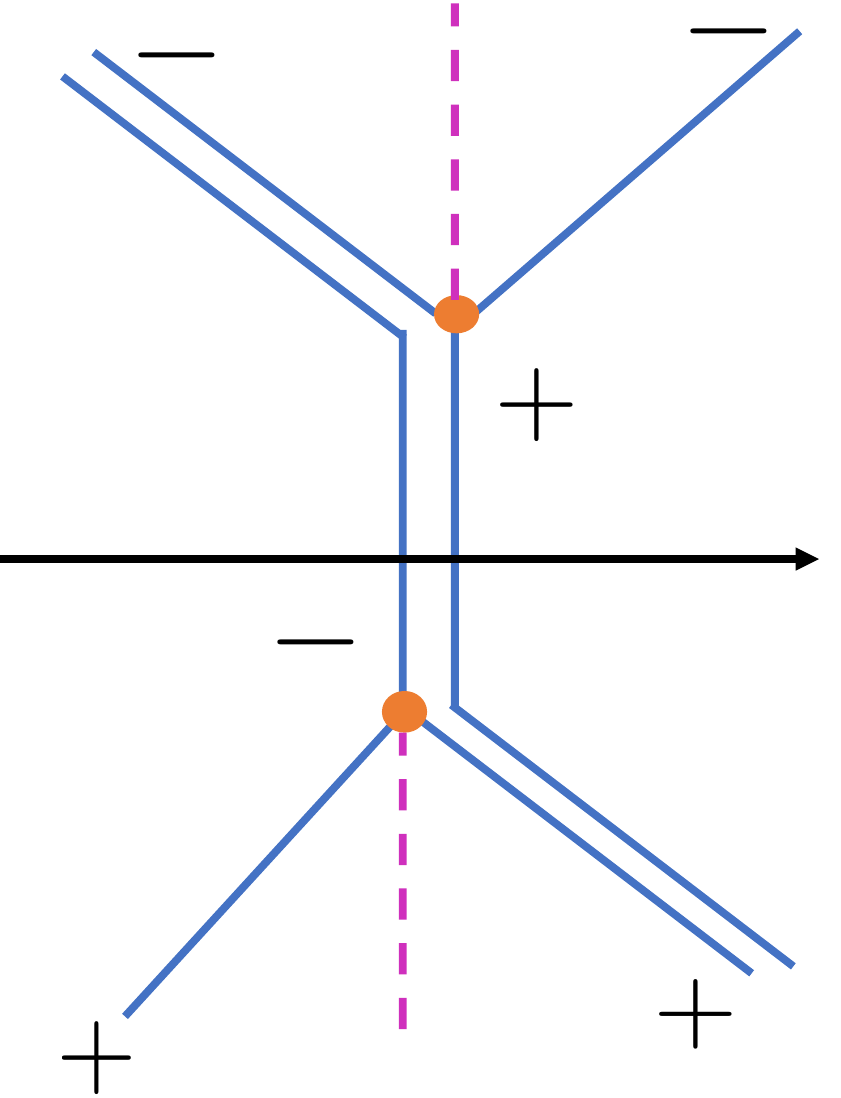}
\caption{\label{fig:nt2} The Stokes line structure of $\omega^2=k^2+v^2t^2$ for $\eta<0$.}
\end{figure}

As the previous example, there appear two turning points but on the imaginary time axis $t=\pm {\rm i}\frac{k}{v}$, from which Stokes segment emanates. We introduce a phase shift for $k$ such that the Stokes segment is decomposed. We illustrate the Stokes line structure in Figs.~\ref{fig:nt1} and \ref{fig:nt2} for each phase shift $k \to k e^{{\rm i}\eta}$ $(\eta \lessgtr 0)$. We denote the turning points as $t=\pm{\rm i} \frac{k}{v}\equiv\pm t_c$. We take the lower limit of the phase integration to be a large negative point $t=t_0<0$, and consider the initial mode function to be
\begin{equation}
    \psi=(t,t_0).
\end{equation}
In the following, we discuss the connection problem of two cases separately.

Let us first discuss the case shown in Fig.~\ref{fig:nt1}. We consider the connection of mode functions from a real negative $t$ to a real positive $t$. We change the integration limit as $\psi=[t_c,t_0](t,t_c)$ and the first Stokes line crossing does not change the form since $(t,t_c)$ is subdominant: $\psi\to \psi'=[t_c,t_0](t,t_c)$. For the second crossing, we change the lower limit as
$\psi'=a b (t,-t_c)$ where $a=[t_c,t_0], b=[-t_c,t_c]$ and we have separated the two factors to specify how we perform the integration. The second crossing leads to the following continuation
\begin{equation}
    \psi'\to\tilde{\psi}=ab\left\{(t,-t_c)+S(-t_c,t)\right\}=(t,t_0)+Sa^2b^2(t_0,t).
\end{equation}
We can decompose $a$ as $a=e^{\frac12( \kappa'+{\rm i}\zeta')}$ where $\kappa'$ and $\zeta'$ are real and these constants are given by $e^{\frac12\kappa'}=[t_c,t_s]$ and $e^{\frac12{\rm i}\zeta'}=[t_s,t_0]$. Here, $t_s$ is the point where the Stokes segment intersects with the real $t$-axis. We also note that $\kappa'>0$ since the Stokes line connecting $t_s$ and $t_c$ is a positive one. We also find that $b=e^{-\kappa'}$. Thus, with the approximation $S\sim {\rm i}$, we find the resultant mode function at a large positive $t$ on the real axis as
\begin{equation}
    \tilde{\psi}=(t,t_0)+{\rm i}e^{-\kappa'+{\rm i}\zeta'}(t_0,t).
\end{equation}

Next, let us discuss the different phase parameter case shown in Fig.~\ref{fig:nt2}. The first Stokes line emanates from $-t_c$, and we take $\psi=[-t_c,t_0](t,-t_c)$, which is analytically continued as
\begin{equation}
    \psi\to\psi'=[-t_c,t_0]\left\{(t,-t_c)+S'(-t_c,t)\right\},
\end{equation}
where $S'$ is a Stokes constant. After change of the lower limit, we finally find the continued mode function in the region on the real $t$-axis with $t>0$ as
\begin{equation}
\tilde{\tilde{\psi}}=[-t_c,t_0]^2S''(t_0,t)+(1+S''S'[-t_c,t_c]^2)(t,t_0).
\end{equation}
Note that the integrals in this expression are rewritten by $\kappa'$ and $\zeta$ as $[-t_c,t_0]^2=e^{-\kappa'+{\rm i}\zeta'}$ and $[-t_c,t_c]=e^{-\kappa'}$ where we have taken into account the signs of Stokes lines. With approximations $S'\sim S''\sim {\rm i}$, we obtain
\begin{equation}
    \tilde{\tilde{\psi}}\sim (1+e^{-2\kappa})(t,t_0)+{\rm i}e^{-\kappa'+{\rm i}\zeta'}(t_0,t).
\end{equation}
Again, we find $\tilde{\psi}\sim\tilde{\tilde{\psi}}$ up to an exponentially small subdominant factor $e^{-2\kappa'}$. Thus, we have confirmed that the different modifications of the Stokes line diagram give us the same result at the leading order. We also note that the normalization of the Bogoliubov coefficient is not satisfied due to the lack of the exponentially suppressed subleading term. Nevertheless, we find this approximation practically useful. Let us note that in our case, the factor $\kappa'$ is given by $2\kappa'=\pi k^2/v^2$, which gives the resultant particle number to be
\begin{equation}
    n=e^{-2\kappa'}=\exp\left(-\frac{\pi k^2}{v^2}\right),
\end{equation}
which precisely reproduces the well known result~\cite{Kofman:1997yn,Kofman:2004yc}.

Finally, we note that as we mentioned in the previous case, the produced number density is actually given by the phase integral along the Stokes segment $[-t_c,t_c]$. This is not so surprising because the Stokes line structure looks the same as the previous toy model if one rotates the complex $t$-plane, or rotate the phase of $\omega$. In various models, one finds the Stokes segments perpendicularly intersecting with the real axis. In such a case, the amount of the particle production would be expressed by the similar integral. Namely, the number density due to such an ``event'' is given by
\begin{equation}
   n_k \sim \exp\left({\rm i}\int_{\bar{t}_c}^{t_c}\omega_k dt\right), \label{eq:Stokesnk}
\end{equation}
where $t_c$ and $\bar{t}_c$ are a pair of the turning points connected by a Stokes segment. In the main text, we use this approximate formula in evaluating the particle production. We should emphasize that this approximation cannot work well when there are poles except at infinity. In such a case, more careful treatment is necessary, and we find such an example in Sec.~\ref{transition} and appendix~\ref{appB}.

\section{Analytic estimation of particle production induced by oscillating background} \label{appC}
\subsection{Rough estimate}\label{appC1}
First, we calculate the Bogoliubov coefficient \eqref{hombuntac}:
\begin{equation}
    |\beta_k| = \exp\left(\int_{t_{c,2n}}^{t_{c,2n+1}} |\Omega_k| dt\right), \label{apptac}
\end{equation}
where $t_{c,2n}$ and $t_{c,2n+1}$ ($n=0,1,2,\cdots$) denote the $(n+1)$-th pair of turning points and the effective frequency $\Omega_k$ is the approximated one \eqref{Omegaapprox2}:
\begin{align}
    \Omega_k^2(t) \approx \frac{k^2}{(Mt/4)^{\frac43}} + m^2 - \left(\xi-\frac14\right)\frac{4M}{t}\sin(Mt) \label{Oapp}
\end{align}
at late time oscillation $Mt \gg 1$. Here, $t_{c,2n}$ ($t_{c,2n+1}$) also corresponds to the start (the end) of $(n+1)$-th tachyonic region. Even with this simplified form, it is not possible to find turning points, and therefore, we consider the following approximation: Since we are now considering the large amplitude case and hence the curvature induced term, more specifically, $\sin(Mt)$ dominates the time dependence of the effective frequency during the integration range. The turning points would be near zeros of $\sin(Mt)$, namely $t = n\pi M^{-1}$, at the leading order. Therefore, we can assume that turning points are $t_{c,n} = (n\pi + \varepsilon_n) M^{-1}$, where $|\varepsilon_n| \ll 1$ and then obtain turning points as
\begin{align}
    t_{c,n} &\approx \left( n\pi \pm \sin^{-1}\Delta_n \right) M^{-1}, \label{tpsR2}
\end{align}
where $\Delta_n$ is the value of the small parameter defined in \eqref{smallD} at $t = n\pi M^{-1}$ and we take the plus (minus) sign when $n$ is even (odd). Stokes segments connect the $2n$-th and $(2n+1)$-th turning points, between which $\phi$ is tachyonic. Now the Bogoliubov coefficient~\eqref{apptac} is approximated as
\begin{align}
    \ln|\beta_{k,n}| &\approx \int_{\sin^{-1}\Delta_{2n}}^{\pi-\sin^{-1}\Delta_{2n+1}} \sqrt{\left( \xi-\frac14 \right) \frac{4M^2}{2n\pi+x}\sin x - (k_p^2 + m^2)} \frac{dx}{M}, \label{tochu1}
\end{align}
where $x \equiv Mt - 2n\pi$. This integral is, however, still too complicated and we cannot perform it analytically because there are too many time dependent terms inside the square root. Here we adopt the following useful method: We ignore all time dependence except $\sin x$, which dominates the time dependence of the effective frequency as mentioned above. Between the two turning points $t_{c,2n}$ and $t_{c,2n+1}$, except for the oscillating part $\sin x$, we fix the time dependence of the effective frequency to be an intermediate value $Mt=(2n+\frac12)\pi$, and now we can calculate the Bogoliubov coefficient \eqref{apptac} analytically:
\begin{align}
    \ln|\beta_{k,n}| &\approx \sqrt{\frac{4\xi-1}{(2n+\frac12)\pi}} \int_{\sin^{-1}\Delta_{2n+\frac12}}^{\pi-\sin^{-1}\Delta_{2n+\frac12}} \sqrt{\sin x - \Delta_{2n+\frac12}} dx.
\end{align}
This integral can be performed analytically and we find
\begin{align}
    \ln|\beta_{k,n}| &= \sqrt{\frac{4\xi-1}{(2n+\frac12)\pi}} \times 4\sqrt{1 - \Delta_{2n+\frac12}} {\rm E}\left( \frac{\cos^{-1}\Delta_{2n+\frac12}}{2} \biggr| \frac{2}{1-\Delta_{2n+\frac12}} \right),
\end{align}
where ${\rm E}(\varphi | k^2)$ is the incomplete elliptic integral of the second kind. Since $\Delta_{2n+\frac12} \ll 1$, we may expand this expression in terms of $\Delta_{2n+\frac12}$ and find
\begin{align}
    \ln|\beta_{k,n}| &= 2\sqrt{\frac{4\xi-1}{(2n+\frac12)\pi}} \left( 2{\rm E}\left( \frac{\pi}{4} \Bigr| 2\right) - {\rm F}\left( \frac{\pi}{4} \Bigr| 2\right) \Delta_{2n+\frac12} \right) + \mathcal{O}\left(\Delta_{2n+\frac12}^2\right), \label{appCtochu}
\end{align}
where ${\rm F}(\varphi | k^2)$ is the incomplete elliptic integral of the first kind. Finally, we obtain the following analytic expression by taking terms up to the leading order in $\Delta_{2n+\frac12}$:
\begin{align}
    n_{k,n} = |\beta_{k,n}|^2 \approx \exp\left[ 4\sqrt{\frac{4\xi-1}{(2n+\frac12)\pi}} \left( 2{\rm E}\left( \frac{\pi}{4} \Bigr| 2\right) - {\rm F}\left( \frac{\pi}{4} \Bigr| 2\right) \Delta_{2n+\frac12} \right) \right].\label{nkana}
\end{align}

As a summary of the derivation of \eqref{nkana}, we mention the essential point of our approximation. In deriving the analytic formula~\eqref{nkana}, we have focused only on the time dependence of the oscillating part $\sin (Mt)$ and ignored the rest. This enables us to perform the integration, while keeping the result as accurate as possible. This idea of an approximation would be useful for other models as well if the time dependence is dominated by a few part of a full effective frequency.

\subsection{Improved estimate}\label{appC2}
\begin{figure}[htbp]
\centering
\includegraphics[width=0.64\textwidth]{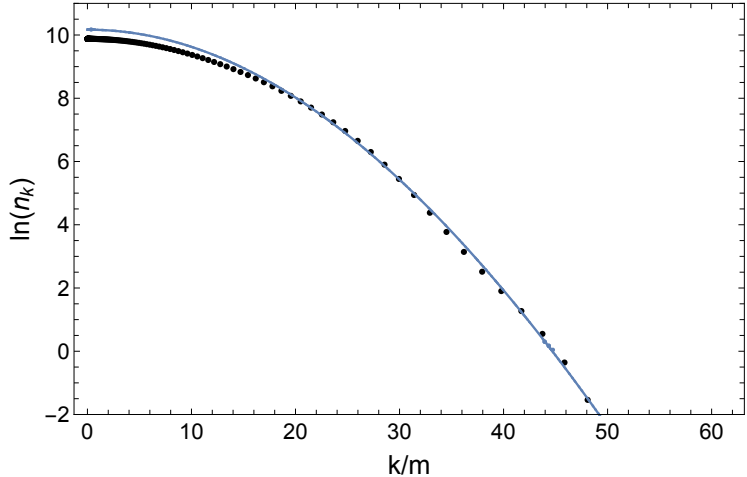}
\caption{\label{fig:improved osc} Comparison between the analytic estimation~\eqref{analytic osc} (blue line) and numerically calculated values (black dots) at the first oscillation $n=0$. For analytic estimation, we use \eqref{appC2tochu}, and for numerical evaluation, we solve the equation of motion with~\eqref{Hosc}, but we used approximated scale factor to be $a=(1+Mt/4)^{2/3}$, which is a good approximation as the solution of~\eqref{Hosc}. We take $M=15m, \xi=10$ in this graph. Note that the first oscillation shows the largest deviation from the numerical result. For larger $n$, the agreement becomes even better. Nevertheless, the agreement for $n=0$ is already excellent.}
\end{figure}

Next, we calculate the Bogoliubov coefficient \eqref{apptac} with the improved approximated effective frequency \eqref{impR}:
\begin{align}
    \Omega_k^2(t) = \frac{k^2}{(1+Mt/4)^{\frac43}} + m^2 - \left(\xi-\frac14\right)\frac{4M^2}{(Mt+4)}\sin(Mt) + \frac{4\xi M^2}{(Mt + 4)^2}.
\end{align}
Since the time dependence of this frequency is even more complicated than the previous one, we again apply the idea used in the previous case, namely we ignore all time dependence except $\sin (Mt)$, which dominates the time dependence of the effective frequency. This approach reduces the problem to the one in the previous case. We take the same steps and find turning points at $Mt_{c,2n} \approx 2n\pi+\epsilon_{2n+\frac12}$ and $Mt_{c,2n+1} \approx (2n+1)\pi-\epsilon_{2n+\frac12}$, where the small quantity $\epsilon_n$ satisfies
\begin{align}
    \sin \epsilon_{2n+\frac12} = \frac{\left(2n+\frac12\right)\pi + 4}{4\xi - 1}\left( \frac{\omega_{k,2n+\frac12}^2}{M^2} + \frac{4\xi}{\left(\left(2n+\frac12\right)\pi + 4\right)^2} \right) \equiv \delta_{2n+\frac12},
\end{align}
where $\omega_{k,2n+\frac12}^2$ is the `bare' energy $k_p^2 + m^2$ at $Mt = \left(2n+\frac12\right)\pi$, namely:
\begin{align}
    \omega_{k,2n+\frac12}^2 = \frac{k^2}{\left( 1+\left(2n+\frac12\right)\pi/4 \right)^{\frac43}} + m^2.
\end{align}
Note that, we have substituted $Mt=\left(2n+\frac12\right)\pi$ except for the oscillating part $\sin Mt$ as explained above.
Solving this equation yields the perturbative quantity $\epsilon_n$. The integral~\eqref{apptac} under our approximation is now written as
\begin{equation}
    \ln\beta_{k,n} \approx \sqrt{\frac{4\xi-1}{4+(2n+\frac12)\pi}}\int_{\epsilon_{2n+\frac12}}^{\pi-\epsilon_{2n+\frac12}}\sqrt{\sin\tau-\sin\epsilon_{2n+\frac12}}d\tau.
\end{equation}
This integral can be performed analytically and we find
\begin{equation}
    \ln\beta_{k,n} = \sqrt{\frac{4\xi-1}{4+(2n+\frac12)\pi}} \times 4\sqrt{ 1-\delta_{2n+\frac12}} {\rm E}\left(\frac{\cos^{-1} \delta_{2n+\frac12}}{2}\biggr|\frac{2}{1-\delta_{2n+\frac12}}\right).
\end{equation}
Since $\delta_{2n+\frac12} \ll 1$, we may expand this expression in terms of $\delta_{2n+\frac12}$ and find
\begin{equation}
    \ln\beta_{k,n} = 2\sqrt{\frac{4\xi-1}{4+(2n+\frac12)\pi}} \left(2E\left(\frac{\pi}{4}\Bigr|2\right) - F\left(\frac{\pi}{4}\Bigr|2\right)\delta_{2n+\frac12}\right)+\mathcal{O}\left( \delta_{2n+\frac12}^2 \right). \label{appC2tochu}
\end{equation}
Finally, we obtain the following estimate by taking terms up to the leading order of $\Delta_{2n+\frac12}$ in \eqref{appC2tochu}:
\begin{align}
    n_{k,n} = |\beta_{k,n}|^2 \approx \exp\left[ 4\sqrt{\frac{4\xi-1}{4+(2n+\frac12)\pi}} \left(2E\left(\frac{\pi}{4}\Bigr|2\right) - F\left(\frac{\pi}{4}\Bigr|2\right)\delta_{2n+\frac12}\right) \right].\label{analytic osc}
\end{align}
We show the comparison between the analytic result and the numerical solution with the exact expression~\eqref{Hosc} in Fig.~\ref{fig:improved osc} for the first oscillation $n=0$, which shows an excellent agreement.

\section{Improvement of particle production rate for a transition model}\label{appB}
\begin{figure}[htbp]
\centering
\includegraphics[width=.50\textwidth]{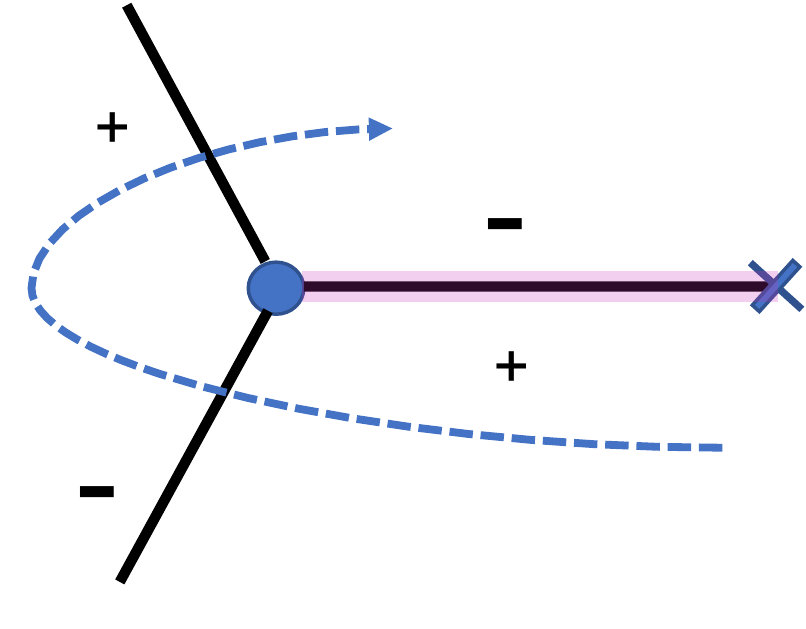}
\caption{\label{fig:pole} A schematic figure of the Stokes line (black solid lines) around one turning point. We consider the connection along the arrow (blue dashed line). The blue dot, blue cross mark and red thick line denote a turning point, a pole and a branch cut, respectively. Solid lines are Stokes lines and the plus and minus signs are that of Stokes lines.}
\end{figure}

In this section, we show how the improved particle number formula~\eqref{improved} is derived. First, we derive a constraint on the Stokes constants by considering the connection around a pole. The schematic figure is shown in Fig.~\ref{fig:pole}.
We consider the mode function in large real positive $t$ region and take the lower limit of a phase integral to be the pole $t_p$, namely our mode function is
\begin{equation}
    f = \frac{A}{\sqrt{\omega_k(t)}}\exp\left(-{\rm i}\int_{t_p}^t \omega_k(t') dt'\right)+\frac{B}{\sqrt{\omega_k}}\exp\left({\rm i}\int_{t_p}^t \omega_k(t') dt'\right)
\end{equation}
where $A$ and $B$ are constants. 
The first Stokes line crossing leads to
\begin{equation}
    f=Ac(t,t_c)+Bc^{-1}(t_c,t)\to f'=Ac(t,t_c)+(Bc^{-1}+S_- Ac)(t_c,t),
\end{equation}
where $c=[t_c,t_p]$ and $S_-$ denotes the Stokes constant, which will be determined later. Note that $c$ is evaluated under the cut and $c>1$. At the second crossing, the mode function is analytically continued to
\begin{align}
    f'&=Ac(t,t_c)+(Bc^{-1}+S_- Ac)(t_c,t)\nonumber\\
    &\to \tilde{f}=(Bc^{-1}+S_-Ac)(t_c,t)+(S_+Bc^{-1}+Ac(1+S_+S_-))(t,t_c).
\end{align}
Finally, taking the lower limit of the integration to be $t_p$, we find
\begin{equation}
    \tilde{f}=(Bc^{-1}+S_-A)(t_p,t)+(S_+B+Ac^{2}(1+S_+S_-))(t,t_p).
\end{equation}
Let us denote $\tilde{f}=A'(t,t_p)+B'(t_p,t)$, and then the relation between $A',B'$ and $A,B$ is written as
\begin{equation}
    \left(\begin{array}{c}A'\\ B'\end{array}\right)=\left(\begin{array}{cc} c^2(1+S_+S_-) & S_+\\
    S_- & c^{-2}\end{array}\right)\left(\begin{array}{c}A\\ B\end{array}\right)\equiv {\cal R} \left(\begin{array}{c}A\\B\end{array}\right).
\end{equation}
This matrix ${\cal R}$ satisfies ${\rm det}\:{\cal R}=1$ and for the case with a first order simple pole, it is known that ${\rm Tr}\:{\cal R}=2$~\cite{Kutlin}. Therefore, we find a constraint 
\begin{equation}
    S_+S_-=-(1-c^{-2})^2.
\end{equation}
Since the effective frequency is real on the real $t$-axis, the following condition holds~\cite{Kutlin},
\begin{equation}
    S_+=-\bar{S}_-.
\end{equation}
This relation leads to $|S_+|^2=|S_-|^2=(1-c^{-2})^2$. Thus, we have fixed the absolute value of the Stokes constant. We note that both $S_+$ and $S_-$ vanish when $c\to1$ and indeed this is the limit of $m\to0$ as we show below.

Next, let us consider the connection problem at the Stokes line crossing. As we show in Fig.~\ref{fig:st_sltower}, there is one Stokes segment that crosses real $t$-axis. If we take into account only this Stokes segment, the connection problem is the same as that discussed in appendix~\ref{appA2}. However, there is one important difference: As we have discussed above, the Stokes constant is no longer ${\rm i}$, but its absolute value is given by $|S_\pm|=|1-c^{-2}|$. Except this point, we can take the same step to obtain the particle number as done in \ref{appA2}. The result is given by\footnote{We have not specified how the Stokes segment is separated into two Stokes lines as is done in Figs.~\ref{fig:nt1} or \ref{fig:nt2}, but it is not important since in either case, the produced negative frequency mode is multiplied by one $S_\pm$. Noting $|S_+|=|S_-|$ we reach the same result for the particle number by taking normalization into account. The phase of $S_\pm$ is not relevant for particle number density either.}
\begin{equation}
    n_k=|\beta_k|^2\sim d^2|S_-|^2= d^2(1-c^{-2})^2,
\end{equation}
where
\begin{equation}
    d = [t_c,\bar{t}_c]=e^{-2\pi k\Delta t},
\end{equation}
and
\begin{equation}
    c = [t_c,t_p]=\exp\left(\pi\Delta t\left(\sqrt{k^2+m^2}-k\right)\right).
\end{equation}
As we have mentioned above, $m\to 0$ limit reads $c\to1$ and the particle number vanishes in this limit as physically expected. Thus, we have derived the improved formula~\eqref{improved}.

\bibliographystyle{JHEP}
\bibliography{ref.bib}

\end{document}